\documentclass{elsart}
      
\usepackage{graphicx}
\usepackage{epsfig,array}
\usepackage{booktabs}
\usepackage[english]{babel}

\newcommand{\beq}{\begin{equation}}
\newcommand{\eeq}{\end{equation}}
\newcommand{\beqa}{\begin{eqnarray}}
\newcommand{\eeqa}{\end{eqnarray}}

\newcommand{\sla}[1]%
        {\kern .25em\raise.18ex\hbox{$/$}\kern-.75em #1}
\newcommand{\mybar}[1]%
        {\kern 0.8pt\overline{\kern -0.8pt#1\kern -0.8pt}\kern 0.8pt}

\begin{document} 
\begin{frontmatter}
\title{Constraints on SUSY Lepton Flavour
 Violation by rare processes}
\author[romeII]{P.Paradisi}
\address[romeII]{University of Rome ``Tor Vergata'' and INFN sez. RomaII, 
	Via della Ricerca Scientifica 1, 
	I-00133 Rome}

\begin{abstract}
We study the constraints on flavour violating terms in low energy SUSY
coming from several processes as  $l_i\rightarrow l_j \gamma$, $l_i\rightarrow l_jl_jl_j$
and $\mu\rightarrow e\ in \ Nuclei$.
We show that a combined analysis of these processes allows us to extract additional
information with respect to an individual analysis of all the processes.
In particular, it makes possible to put bounds 
on sectors previously unconstrained by $l_i\rightarrow l_j \gamma$.
We perform the analysis both in the mass eigenstate and in the mass insertion approximations 
clarifying the limit of applicability of these approximations. 

\end{abstract}

\end{frontmatter}

\section{Introduction}
\label{sec:introduction}
Neutrino oscillation experiments have established the existence of lepton family number 
violation processes.
So, as a natural consequence of neutrino oscillations, one would expect flavour mixing
in the charged lepton sector.
This mixing can be manifested in rare decay processes such as $\mu\rightarrow e\gamma$,
$\tau\rightarrow \mu\gamma$, etc. However, if only the lepton yukawa
couplings carry this information on flavour mixing, 
as in the Standard Model with massive neutrinos, the expected 
rates of these processes are extremely tiny being proportional to the
ratio of masses of neutrinos over the masses of the W bosons \cite{SMmueg}.
These values are very far from the present and upcoming experimental upper bounds 
that we can read in Table 1. \\
\begingroup
\begin{table}
\label{tb2}
\begin{tabular}{cccc}
\toprule   
&Process & Present Bounds & Expected Future Bounds  \\[0.2pt] 
\toprule
(1) &  BR($\mu \to e,\gamma$) & $1.1~ \times~ 10^{-11}$  &
$\mathcal{O}(10^{-13} - 10^{-14})$ \\ 
(2) &  BR($\mu \to e,e,e$ ) & $1.1~ \times~ 10^{-12}$ &
$\mathcal{O}(10^{-13} - 10^{-14})$ \\ 
(3) &  BR($\mu \to e$ in Nuclei) & $1.1~ \times~ 10^{-12}$ &
$\mathcal{O}(10^{-13} - 10^{-14})$ \\ 
(4) &  BR($\tau \to e,\gamma$) & $3.1~ \times~ 10^{-7}$ &
$\mathcal{O}(10^{-8}) $ \\ 
(5) &  BR($\tau \to e,e,e$) & $2.7~ \times~ 10^{-7}$ &
$\mathcal{O}(10^{-8}) $ \\ 
(7) &  BR($\tau \to \mu,\gamma$) & $6.8~ \times~ 10^{-8}$ &
$\mathcal{O}(10^{-8}) $ \\ 
(8) &  BR($\tau \to \mu, \mu, \mu$) & $2~ \times~ 10^{-7}$ &
$\mathcal{O}(10^{-8}) $ \\ 
 
\toprule   
\end{tabular}
\caption{Present and Upcoming experimental limits on various leptonic processes 
\cite{mega,belletmg,belle2,babar,belletalk,psi}}
\end{table}
\endgroup
In a supersymmetric (SUSY) framework the situation is completely different \cite{vempatirew}.
For instance, the supersymmetric extension of the see-saw model \cite{seesawrefs}
provides new direct sources of flavour violation, namely the possible
presence of off-diagonal soft terms in the slepton mass matrices and 
in the trilinear couplings \cite{fbam}. In practice, flavour violation would 
originate from any misalignment between fermion and sfermion mass eigenstates. \\
One of the major problems of low energy supersymmetry is to understand why all LFV 
processes are so suppressed. This suppression imposes very severe 
constraints on the pattern of the sfermion mass matrices which must be either very 
close to the unit matrix in the flavour space (flavour universality) or almost 
proportional to the corresponding fermion mass matrices (alignment).  
Both universality and alignment can be either present as a kind of ``initial'' 
conditions or as a result from some dynamics of the theory. \\ Given a specific SUSY model, 
it is possible, in that context, to make a full computation of all the FCNC
(and possibly also CP violating) phenomena. This is the case, 
for instance, of the constrained minimal supersymmetric
standard model (CMSSM) where these detailed computations led to the result of utmost
importance that this model succeeds to pass unscathed all the severe
FCNC and CP tests. However, given the variety of 
options that exists in extending the MSSM (for instance embedding it in some more 
fundamental theory at larger scale), it is important to have a way to extract from 
the whole host of FCNC and CP phenomena a set of upper limits on quantities which 
can be readily computed in any chosen SUSY frame. Namely, one needs some kind of 
model-independent parameterization of the FCNC and CP quantities in SUSY to have 
an accessible immediate test of variants of the MSSM. \\
The best parameterization of this kind that has been proposed is in the framework 
of the so-called mass insertion approximation (MI) \cite{hkr}.
One chooses a basis for the fermion and sfermion states where all the couplings 
of these particles  with neutralinos or with gluinos are flavour diagonal, 
while the flavour changing is exhibited by the non-diagonality of the sfermion propagators. 
Denoting with $\Delta_{ij}$ the off-diagonal terms (in flavour space) in the sfermion 
mass matrices, the sfermion propagators can be expanded as a series in terms of 
$\delta_{ij}={\Delta_{ij}}/{\tilde m^ 2}$ where $\tilde m$ is an
average sfermion mass squared.
As long as $\delta_{ij}$ is much smaller than one, we can just take the first term 
of this expansion and then the experimental information concerning FCNC and 
CP violating phenomena translates into upper bounds on these $\delta_{ij}$. \\
Obviously, the above mass insertion method is advantageous because one does not need 
the full diagonalization of the sfermion mass matrices to perform a test of the considered 
susy model in the FCNC sector.\\ 
Pioneering studies \cite{amgab,gabbiani} 
considered only the photinos and gluinos me\-diated FCNC contributions
because a complete inclusion of the neutralino and chargino sector contributions
would require a complete specification of the model. 
On the contrary, the spirit of their study was to provide a model-independent way 
to make a first check of the FCNC impact on classes of SUSY theories. \\
Subsequently, general studies in the leptonic sector \cite{MS} included the
chargino and neutralino contributions using a generalization of the MI,
that we call GMI \cite{prs}, which consists in an approximation where 
the gaugino-higgsino mixings are also treated as insertions in the
propagators of the charginos and neutralinos inside the loop. 
With this additional insertion approximation, it is not necessary to
fix a particular scenario to analyze the dependence on the many mass
parameters. However, in the lepton sector,
interferences among amplitudes are generically present in the models and they would affect 
the limits on the $\delta_{ij}$. This happens, for instance, for $\delta^{RR}$ due to 
a destructive interference between the bino and bino-higgsino amplitudes.
Moreover $\tau\rightarrow \mu\gamma$  and $\tau\rightarrow e\gamma$  experimental bounds are 
not so stringent as in $\mu\rightarrow e\gamma$ so that we can have not so tiny 
insertions independently from cancellations. \\
In this context, the target of this work is to study all the possible constraints on flavour
violating terms in low energy SUSY coming from several processes as 
$l_{i}\rightarrow  l_{j} \gamma$, $l_{i}\rightarrow  l_{j}l_{j}l_{j}$
and $\mu\rightarrow e\ in\ Nuclei$ and, subsequently, to clarify which
is the limit of applicability of the MI and of its generalization.\\ 
The approach is to fix a specific model of known spectrum (CMSSM) and
to compare, systematically, the predictions of the full computation
\cite{hisano} with respect to the approximate, MI and GMI, computations \cite{HN,MS}.\\ 
In particular, we focused on the RR sector where strong cancellations
make the sector unconstrained. 
We show that these cancellations prevent us from getting a bound in
the $RR$ sector both at the present and also in the future when the
experimental sensitivity on $Br(l_i \rightarrow l_j \gamma)$ will be increased.
So, being our aim to find constraints in the $RR$ sector, we examined other LFV processes as
$l_{i}\rightarrow  l_{j}l_{j}l_{j}$ and $\mu\rightarrow e\ in\ Nuclei$. \\
Finally we took into consideration the bounds on the various double mass insertions
$\delta_{23}^{AB} \delta_{31}^{CD}$, with $A, B, C, D=L, R$.
These limits are important in order to get the largest amount of information 
on SUSY flavour symmetry breaking.
                                                                                                        

\section{LEPTON FLAVOUR VIOLATION IN SUPERSYMMETRY }

The study of lepton flavour violation in SUSY scenarios is one of the most 
promising subjects in low energy supersymmetric phenomenology, in that
it shows, if observed, a clear signal of physics beyond the Standard Model \cite{profumo,cannoni}. \\
The source of LFV is the soft supersymmetry breaking lagrangian which,   
in general, contains a too large number of FV couplings, so that 
the predicted branching ratios for LFV processes usually exceed
phenomenological constraints \cite{fcncreview} :
this is a typical example of the SUSY flavour-problem. \\
The usual way to solve this problem is to consider Lagrangians
which result from models that break SUSY in a flavour blind manner, as in mSUGRA or 
Anomaly mediated supersymmetry breaking (AMSB), etc \cite{kanekingphyrep}. \\
Even in this case, in general, flavour is violated in the Lagrangian at the weak scale.
For instance, LFV can be induced by the existence of new particles at high scale 
with flavour violating couplings to the SM leptons
(as right handed neutrinos in a see-saw model \cite{amgab}) or the
presence of new Yukawa interactions, as in superGUTs theories
\cite{bhs,so10}. In this last case the flavour violation 
is communicated to low energy fields through renormalization effects. \\
On the other hand, in models based on supergravity or superstring theories,
nonuniversal soft terms are generically present in the high scale effective \\ 
Lagrangian 
\cite{brignoleibanez}. \\ 
In models with flavour symmetry imposed by a Froggatt-Nielsen mechanism, 
flavour violating corrections to the soft potential could be large 
\cite{dudassavoy,oscarross1,oscarross2}. \\
Irrespective to the source of these FV entries, our approach is to
assume LFV at low energy and to try to bound the FV terms present in slepton mass matrices. \\
The processes we are going to study are
$l_{i}\rightarrow  l_{j} \gamma$, $l_{i}\rightarrow  l_{j}l_{j}l_{j}$ and 
$\mu\rightarrow e\ in\ Nuclei$ that are mediated, at one loop level,
by neutralinos, charginos and sleptons. The relevant interaction
Lagrangian for the considered processes is:

\beqa
\mathcal{ L}=\overline{l}_i\left(C^{R}_{iAX}P_R+C^{L}_{iAX}P_L\right){\tilde
\chi}^{-}_{A}\ {\tilde\nu}_X+\overline{l}_i\left(N^{R}_{iAX}P_R+N^{L}_{iAX}P_L\right)
{\tilde\chi}^0_A{\tilde l}_X.
\eeqa

We can always work in the basis where the charged lepton masses 
and the gauge couplings are flavour diagonal.
In this basis, in general, the slepton mass matrix is not diagonal and its
off-diagonal entries induce the LFV. \\
Our aim is to use the couplings $C_{iAX}^{R,L}$ and $N_{iAX}^{R,L}$, which are functions
of the FV entries, to constrain the structure of the slepton mass matrix. 

\section{\bf MASS INSERTION APPROXIMATION}
In the spirit of the mass insertion approximation,
we treat the off-diagonal elements of the slepton mass matrix as insertions.
Our convention for the slepton mass matrices is:
$$(\tilde{l}^{\dag}_L \tilde{l}^{\dag}_{R}) 
  \left(\begin{array}{ccc}
    {m_L^2(1+\delta_{LL})} & {(A-\mu\tan\beta)m_{l}+m_Lm_R\delta_{LR}} \nonumber \\
    {(A-\mu\tan\beta)m_{l}+m_Lm_R\delta_{LR}}^{\dag} & {m_R^2(1+\delta_{RR})} 
  \end{array} \right) \nonumber
  \left(  \begin{array}{ccc}
        {\tilde{l} _L} \nonumber \\
        {\tilde{l} _R} \nonumber
\end{array} \right) \nonumber$$\\
where $m_L$ and $m_R$ are, respectively, the average masses  of the L and R sleptons 
and $A=am_0$ with $a\simeq O(1)$. \\
The MI now corresponds to a development of the slepton propagators 
around the diagonal with the average slepton masses, $m_L^2$ and $m_R^2$. 
In practice, we are working in the basis of leptons, neutralinos and charginos 
mass eigenstates and slepton weak eigenstates. \\
The FV is parametrized through a flavour violating mass insertion 
in the virtual slepton line. 
In this basis, the gaugino couplings are flavour-diagonal and so the Lagrangian in eq.1 
takes the form: 
\beqa
\mathcal{-L}&=&
\tilde{l}_{Li}^{\dag}\overline{\tilde{\chi}_{A}^{0}} \bigg({N_{LR}^{A(i)}}
P_{R}+{N_{LL}^{A(i)}}P_{L}\bigg)l_{i}+\tilde{l}_{Ri}^{\dag}
\overline{\tilde{\chi}_{A}^{0}}\bigg({N_{RR}^{A(i)}} P_{R}+{N_{RL}^{A(i)}}P_{L}\bigg)l_{i}
\nonumber\\
&+&
\tilde{\nu}_{i}^{\dag}
\overline{\tilde{\chi}_{A}^{-}} \bigg({C_{LR}^{A(i)}}P_{R}+{C_{LL}^{A(i)}}P_{L}\bigg)l_{i}+h.c.,
\,\,\,\,\,\,\,\,\,\,\,i=e,\mu,\tau
\eeqa
where the coefficient $C_{B,C}^{A(i)}$ and $N_{B,C}^{A(i)}$ (with $B,C=(L,R)$) are:
\beqa
C_{LL}^{A(i)}&=&g_{2}(O_{R})_{A1}\,\,,\,\,\,\,\,\,\,
C_{LR}^{A(i)}=-\frac{g_{1}}{\sqrt2}\frac{m_{l_i}} {M_Wc_{\beta}}(O_{L})_{A2}\nonumber\\
N_{LL}^{A(i)}&=&-\frac{g_{2}}{\sqrt2}(O_{N})_{A2}-\frac{g_{1}}{\sqrt2}(O_{N})_{A1}\,\,,\,\,\,\,\,\,\,
N_{RR}^{A(i)}=\sqrt2 g_{1}(O_{N})_{A1}\nonumber\\
N_{LR}^{A(i)}&=&N_{RL}^{A(i)}=\frac{g_{1}}{\sqrt2}\frac{m_{l_i}}{M_Wc_{\beta}}(O_{N})_{A3},
\eeqa
where $O_{L,R}$ and $O_{N}$ are hermitian matrices that diagonalize 
the chargino and neutralino mass matrices, respectively.

\subsection{\bf Neutralino-chargino sector: GMI approximation.}

In the slepton mass matrix, we were able to factorize the source of LFV, 
namely the mass insertions $\delta_{ij}$.\\
We would like to have a similar tool in the chargino-neutralino sector too. 
In fact, it is difficult to understand the dependence of physical quantities 
on the elements of the chargino-neutralino mass matrices. 
For instance, in the GMI approximation we treat both the off-diagonal elements 
(flavour violating or not) of the slepton mass matrices and the
off-diagonal terms (flavour conserving) of the chargino and neutralino
mass matrices as mass insertions. \\
In order to understand the validity of this other kind of approximation
let us consider the chargino mass matrix $M_C$: 

$$ (\tilde{\overline{W}}_{R}^{-} \ \ \tilde{\overline{H}}_{2R}^{-}) 
  \left( \begin{array}{ccc}
    {M_2} & {\sqrt{2}m_W \cos\beta} \\
    {\sqrt{2}m_W \sin\beta} & {\mu} 
  \end{array} \right) 
  \left(  \begin{array}{ccc}
        {\tilde{W}^- _L} \\
        {\tilde{H}^- _{1L}}
  \end{array} \right)
  + h.c.,$$\\
diagonalized by $2\times2$ hermitian matrices $O_L$ and $O_R$ 
as $O_RM_CO_L^T=(M_C)_{diag}$. Now, assuming that 
$M_{2,1}$, $\mu$, $|M_{2,1}\pm\mu|\simeq\Lambda>\!\!>M_Z$,
we obtain as approximate mass eigenstates 
$\tilde{W}^-$ and $\tilde{H}^-$ with masses $M_2$ and $\mu$, respectively.
Actually, corrections to the spectrum start from the order $O(M_Z^2/\Lambda^2)$, 
while the off diagonal elements of the $O_R, O_L$ matrices are of the form  
$M_W[M_2 s_{\beta}(c_{\beta})+\mu c_{\beta}(s_{\beta})]/[M^2_2-\mu^2]$ where 
$s_{\beta}\!=\!\sin\beta$ and $c_{\beta}\!=\!\cos\beta$.
The approximations made follow naturally from scenarios like m-SUGRA
in which a large $\mu$ term is required in order to get correct 
electroweak symmetry breaking (see for instance table II in \cite{bartl}).
The neutralino spectrum can be analyzed in a very similar way. It turns out that 
a ``bino-like'' LSP can very easily have the right cosmological abundance to make 
a good dark matter candidate, 
so from this point of view the large $\mu$ limit may be preferred.\\ 
In this approximation the amplitudes of the analyzed processes have 
two kinds of contributions: one without off-diagonal MI in the
gauginos propagator, i.e. pure $\tilde{B}$ and $\tilde{W}$, the other with one MI mixing
$\tilde{B}\!-\!\tilde{H}$, $\tilde{W}\!-\!\tilde{H}$, proportional to $M_W$.
The approximation consists in keeping at most one flip insertion, only the terms of 
$\mathcal O(M_W/M_i, M_W/\mu)$, so we are neglecting terms of order 
$\mathcal O(M_W^2/m^2_{susy})$. 
In this basis both charginos, neutralinos and sleptons are in the flavour-eigenstates
and the expressions of the amplitudes are very easy to understand.  

\section{mSUGRA spectrum}

The aim of our analyses is to bound all the leptonic $\delta_{ij}$'s and to establish 
the limit of applicability of the approximate methods presented above.\\
In this context, we prefer to work in a well defined model (m-SUGRA)
to reduce the number of free parameters and to simplify the
physical interpretation of the results.\\ 
Let us recall the constraints arising in mSUGRA, where the universality 
assumption reduces the parameters at the Planck scale to a common scalar mass, $m_0$, 
a common gaugino mass, $M_{1/2}$, and a universal $A=am_0$ term with $a\simeq O(1)$.\\
At the low energy scale, after the RGE running, 
the parameters $M_1$, $M_2$, $m_L$ and  $m_R$ are obtained as follows:    
\beqa
M_{i}(m_W)&\simeq& \frac{\alpha_i(m_W)}{\alpha_i(M_U)}M_{i}(M_U)\nonumber\\\nonumber\\
m^2_L(m_W)&\simeq&  m_{L}^2(M_U) +0.5 M_{2}^2(M_U) +0.04 M_{1}^2(M_U)\nonumber\\\nonumber\\
m^2_R(m_W)&\simeq&  m_{R}^2(M_U) +0.15M_{1}^2(M_U)\nonumber
\eeqa
where $M_U$ is the unification scale and the mSUGRA constraints
are satisfied when $M_{1}(M_U) = M_{2}(M_U) = M_{1/2}$ and 
$m^2_R(M_U) = m^2_L(M_U) = m^2_0$.\\   
A very important constraint in mSUGRA comes from the radiative
electroweak breaking condition that requires a fine-tuned $\mu$ in
order to fulfill the minimum condition:
\beqa
|\mu^2|+\frac{M_{Z}^2}{2}\simeq m_{0}^2 \,\frac{1+0.5
\tan^2\beta}{\tan^2\beta-1}+M_{1/2}^2 \,\frac{0.5+3.5
\tan^2\beta}{\tan^2\beta-1}\ ,\nonumber
\eeqa
so, the spectrum of the model is completely known given the following set of parameters:
$m_0$, $M_{1/2}$, $a$, $\tan\beta$, $|\mu|$. 

\section{LFV processes: $l_{i}\rightarrow l_{j}\gamma$ , $l_{i}\rightarrow 3 l_{j}$ , 
$\mu\rightarrow e \ in \ Ti$. }

Let us discuss the branching ratios of the LFV rare processes as
$l_{i}\rightarrow  l_{j}\gamma$.
The amplitudes of the processes take a form

\beqa
T=m_{l_i}\epsilon^{\lambda}\overline{u}_j(p-q)[iq^\nu\sigma_{\lambda\nu}(A_{L}P_{L}
+A_{R}P_{R})]u_i(p)\nonumber
\eeqa
where $p$ and $q$ are momenta of the leptons $l_k$ and of the photon respectively.\\
The chirality flip of this transition is the reason of the appearance of the 
$m_{l_i}$ factor in the operator. The branching ratio of $l_{i}\rightarrow l_{j}\gamma$ is given by 
\beqa
\frac{BR(l_{i}\rightarrow  l_{j}\gamma)}{BR(l_{i}\rightarrow  l_{j}\nu_i\bar{\nu_j})} = 
\frac{48\pi^{3}\alpha}{G_{F}^{2}}(|A_L^{ij}|^2+|A_R^{ij}|^2)
\ \sim  \ \frac{\alpha^3}{G_{F}^{2}}\frac{\delta_{ij}^2}{\tilde m^{4}} \tan\beta^2\nonumber
\eeqa
with $\tilde m$ a typical susy scale. 
In SUSY, this chirality flip can be implemented in the external fermion line or at Yukawa 
vertex or in the internal sfermion line through a flavour conserving FC mass insertion
$\Delta^{RL}_{ii} = (A-\mu\tan\beta)m_{i}$.\\ 
Each  coefficient in the above formula can be written as a sum of two terms,
$$A_{L,R}=A_{L,R}^{n}+A_{L,R}^{c}$$ where $A_{L,R}^{n}$ and $A_{L,R}^{c}$ 
stand for the contributions from the neutralino loops and from the chargino 
loops respectively.\\ 
Finally, we consider the $l_{i}\rightarrow  3l_{j}$ and $\mu\rightarrow e\ in\ Ti$ processes.
Both of them get contributions from penguin-type diagrams
(with photon or Z boson exchanges) and from box-type diagrams. However the $\gamma$ 
penguin-type contribution is enhanced by a $\tan\beta$ factor with respect to 
the other contributions so it dominates and one can find the simple theoretical relations:
$$
Br(l_{i}\rightarrow l_{j}l_{j}l_{j})\ \simeq 
\ \ 7\times10^{-3}BR(l_{i}\rightarrow  l_{j}\gamma),
$$
$$
Br(\mu\rightarrow e \ in \ Ti) \ \simeq \ \ 6\times10^{-3}BR(\mu \rightarrow e \gamma).
$$
Now, imposing the actual experimental upper limits on the above LFV processes, 
we get the upper bounds on the $\delta_{ij}$'s from each process. In particular,
the ratios among the $\delta_{ij}$'s upper bounds from different processes are: 
$$\delta_{ij}^{l_i \rightarrow 3l_j}\ \simeq \ 4\ \delta_{ij}^{l_i \rightarrow l_j \gamma}$$
$$\delta_{21}^{\mu\rightarrow e\ in\ Ti}\ \simeq \ 4\ \delta_{21}^{\mu\rightarrow e \gamma}.$$
We note that $l_i \rightarrow l_j \gamma$ give the more stringent bounds on the $\delta_{ij}$'s.
However, these relations are no longer true in special regions, where, strong cancellations 
reduce $Br(l_i \rightarrow l_j \gamma)$ by several order of magnitude and
destroy the above relations (see the discussion about $\delta_{RR}$ bounds
in the next section).\\ 
In this paper, we will consider only the contributions from
neutralino and chargino sectors. However, Higgs bosons ($h^0,H^0,A^0$) are also 
sensitive to flavour violation and mediate processes such as 
$\mu \rightarrow e\,\,in\,\,Nuclei$ \cite{okada}, $\tau \to 3 \mu $ \cite{t3mu} or
$\tau \to \mu \eta$ \cite{sher} or $\tau \to \mu(e)\gamma$ \cite{higgsmio}. 
The amplitudes of these processes are sensitive to a 
higher degree in $\tan\beta$ than the chargino/neutralino ones (the BRs grow as
$(\tan\beta)^6$, though they are suppressed by additional Yukawa couplings) and thus
could lead to large branching fractions at large $\tan\beta$ \cite{andrea}.

\section{Bounds on $\delta^{ij}$ from $l_i \rightarrow l_j \gamma$ ,
$l_{i}\rightarrow 3 l_{j}$ , $\mu\rightarrow e \,\, in \,\,Nuclei$ }
The approach we follow to put bounds on the various $\delta^{ij}$
is to consider only one mass insertion contributing at a time for the 
$Br(l_i\rightarrow l_j\gamma)$.
Each off-diagonal FV entry in the slepton mass matrices is put equal to zero, 
except for the FV insertion we are interested to constrain.\\
Now, what we mean by full computation is to diagonalize the slepton mass matrix
(with only one $\delta^{ij}$ term) and to use the experimental
$Br(l_i\rightarrow l_j\gamma)$ limits to impose constraints on each insertion type.\\  
Imposing that the contribution of each $\delta^{ij}$ does not exceed (in absolute value)
the experimental bounds, we obtain the limits on the $\delta^{ij}$'s, barring
accidental cancellations. 
\footnote{In SUSY GUT theories there are possible correlations between 
the hadronic and leptonic FV effects
\cite{ourprl,workinprogress}.}.\\ 
Hence forward, to be as clear as possible, we will speak in the MI language.

\subsection{\bf Bounds on $\delta_{LL}$}

The amplitudes proportional to $\delta_{LL}$ receive both $U(1)$ and $SU(2)$ type contributions.
In the first case we can have a pure $\tilde B$ exchange, with chirality-flip implemented
in the external (internal) fermion (sfermion) line or at yukawa vertex
from $\tilde B-\tilde H^0$.
In the second case we have $\tilde W$ and $\tilde W-\tilde H$ exchange
both for charginos and for neutralinos. However in the $\tilde W$ case, 
because the $SU(2)$ Gauge fields don't couple to right-handed fields,      
we can't make a chirality flip in the internal sfermion line.
When the chirality is flipped in the external lepton line, the amplitude 
does not contain the $\mu$ mass term and is not $\tan\beta$ enhanced.\\ 
In the following we give the expression for the amplitude 
$A^{21}_{L2}=(A^{21}_{L2})_{SU(2)}+(A^{21}_{L2})_{U(1)}$ of the 
$l_i\rightarrow l_j\gamma$ processes in the GMI approximation:
\beqa
\left(A^{ij}_{L}\right)_{SU(2)}\!\!\!=\!\!\frac{\alpha_{2}}{4\pi}\Delta_{LL}^{ij}
\bigg[\frac{f_{1n}(a_L)\!+\!f_{1c}(a_L)}{m^4_{L}}\!+\!
\frac{\mu M_{2}t_{\beta} }{(M_{2}^2\!-\!\mu^2)}
\frac{\left(f_{2n}(a_L,b_L)\!+\!f_{2c}(a_L,b_L)\right)}{m^4_{L}}\bigg]\nonumber
\eeqa
\beqa
\left(A^{ij}_L\right)_{U(1)}=\frac{\alpha_{1}}{4\pi}\Delta_{LL}^{ij}\,\,
&\bigg[&\frac{f_{1n}(a_L)}{m^4_{L}}+
\mu M_{1}t_{\beta}\bigg(\frac{-f_{2n}(a_L,b_L)}{m^4_{L}(M_{1}^2\!-\!\mu^2)}+
\frac{1}{(m^{2}_{R}-m^{2}_{L})}\nonumber\\\nonumber\\
&\cdot&
\bigg(\frac{2f_{2n}(a_L)}{m^4_{L}}\!+\!\frac{1}{(m^{2}_{R}\!-\!m^{2}_{L})}
\bigg(\frac{f_{3n}(a_R)}{m^2_{R}}\!-\!\frac{f_{3n}(a_L)}{m^2_{L}}\bigg)\!\bigg)\!\bigg)\!\bigg],
\nonumber
\eeqa
where $a_{L}=M^{2}_{1,2}/m^{2}_{L}$ for $U(1)$ or $SU(2)$ contributions, respectively,
while $a_{R}=M^{2}_{1}/m^{2}_{R}$ and $b_{L,R}=\mu^2/m^{2}_{L,R}$.
The loop functions $f_{i(c,n)}(x)$'s are reported in appendix A 
while $f_{i(c,n)}(x,y)=f_{i(c,n)}(x)-f_{i(c,n)}(y)$ and $t_{\beta}\!=\!\tan\beta$.\\
In the above equations, for completeness, we included the subdominant contributions that are not 
$\tan\beta$ enhanced.\\
In fig.1 we can see the bounds on  $\delta_{21}^{LL}$ and
$\delta_{32}^{LL}$ coming from $\mu\rightarrow e \gamma$ and $\tau\rightarrow \mu \gamma$, 
respectively. 
As for $\tau\rightarrow e\gamma$, the limit on $\delta_{31}^{LL}$ is such that 
$\delta_{31}^{LL}/\delta_{32}^{LL}= (Br(\tau\rightarrow e\gamma)/Br(\tau\rightarrow \mu\gamma))^{1/2}$.
The red lines correspond to the full computations while the green and the blue
lines refer to the mass insertion (MI) and to the generalized mass insertion (GMI)
approximations, respectively.\\  
In the $\delta_{21}^{LL}$ case, MI and full computations give indistinguishable results
and the GMI gives a very satisfactory approximation (see the next section for a quantitative
estimate). In the $\delta_{32}^{LL}$ case the degree of approximation is worse than in the
$\delta_{21}^{LL}$ case. The motivation is that the flavour violating (FV) and conserving (FC) 
insertions of the $32$ sector are generally larger than those relative
to the $21$ sector. For instance, $\delta_{32}^{LL}\simeq 10^2\delta_{21}^{LL}$ and
$\delta_{33}^{RL}= m_{\tau}/m_{\mu} \delta_{22}^{RL}$.\\     
A large FV insertion, as for $\delta_{32}^{LL}= 0.3$, produces a sizable distinction 
between approximated and full computations as it must be when we go away from 
the perturbative region.\\ The effect of a large FC insertion is evident 
for small $m_0$, where the $\mu$ term is much larger than the slepton masses,
so that to treat $m_{\tau}\mu\tan\beta /\tilde m^2_{L,R}$ as insertion is not 
properly correct.\\
In fig. 1, the most appreciable deviations between the full computation 
and the GMI one are due to the approximations in the neutralino mixing
and they tend to vanish in the large $\tan\beta$ limit,
where the neutralino (chargino) mass matrix expansion is better justified.\\
We remark that we don't see these deviations 
in the regions with chargino or pure-Bino dominance (in fact these last 
contributions are lesser affected by the GMI approximations).
The bounds are rather insensitive to the sign of $\mu$.

\begin{figure}[ht]
\begin{tabular}{cc}
\includegraphics[scale=0.35]{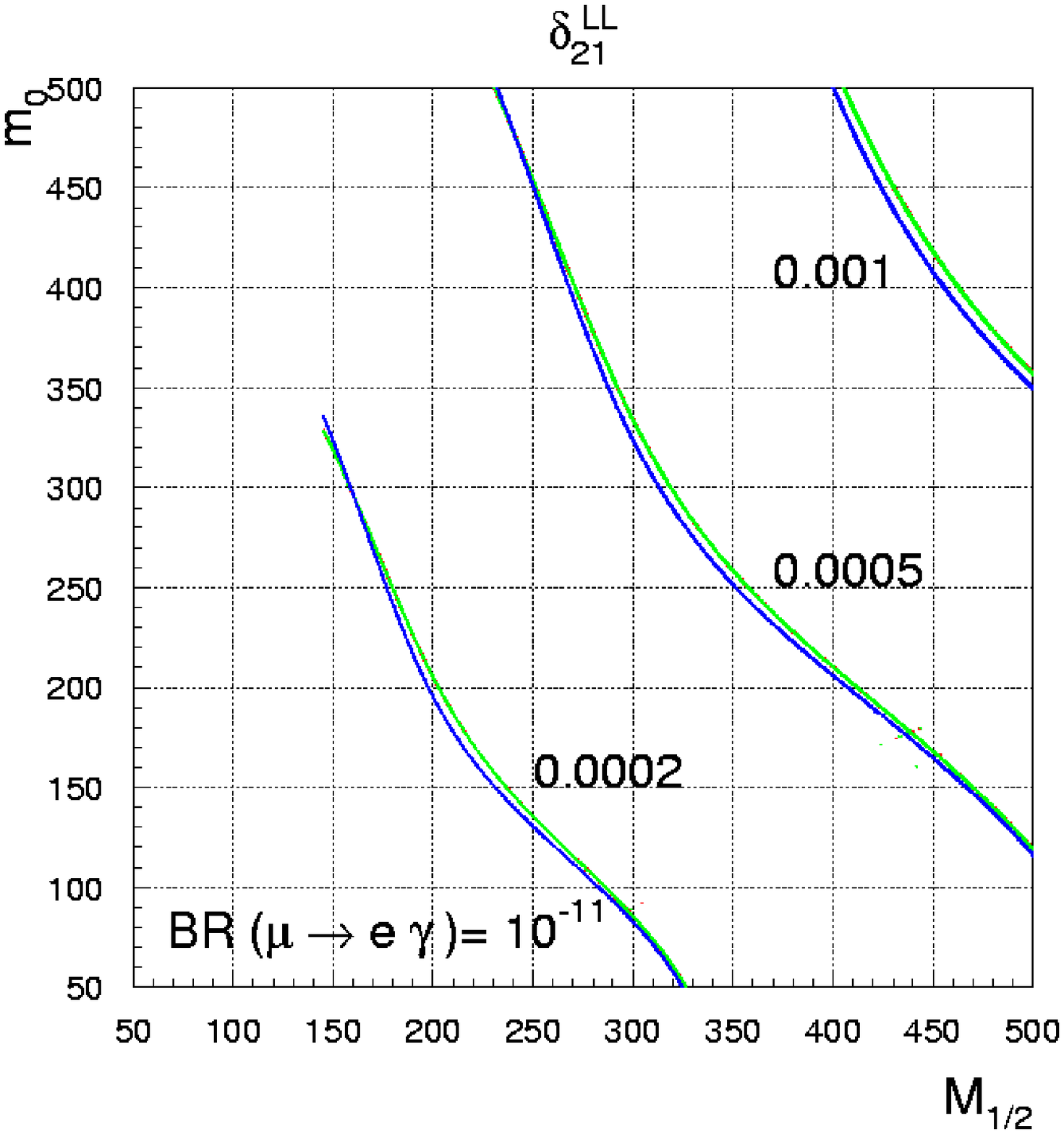} &
\includegraphics[scale=0.35]{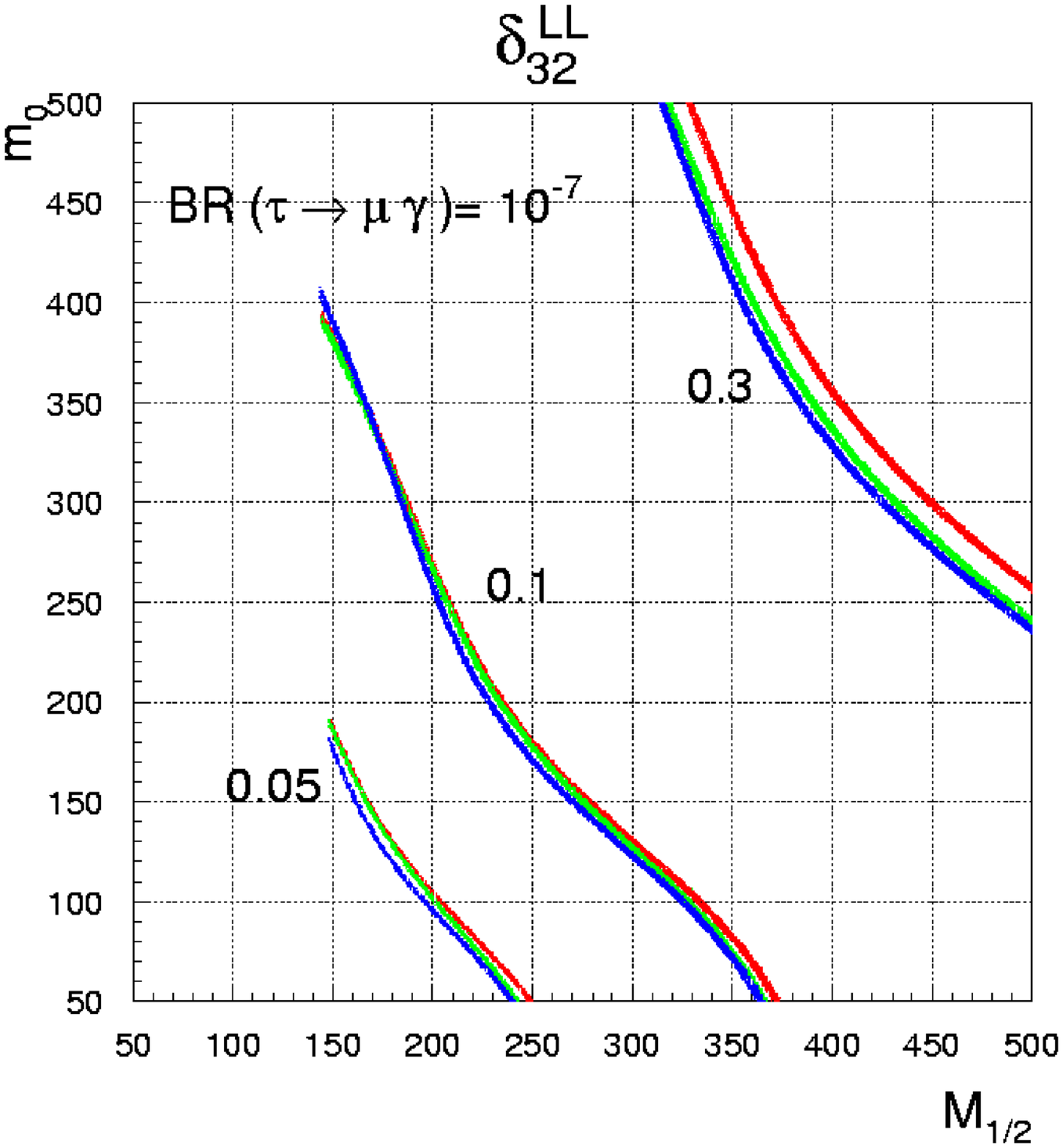}
\end{tabular}
\caption{Upper limits on $\delta^{ij}_{LL}$ from $l_i\rightarrow l_j\gamma$.
In the plots we set $\mu>0$, $\tan\beta=10$ and a=0. 
Red lines correspond to full computation,
green and blue lines to MI and GMI approximations respectively. }
\end{figure}

\subsection{\bf Bounds on $\delta_{LR}$}

In this case, the only contribution arises from the $\tilde B$ exchange
and the amplitude does not contain a $\tan\beta$ factor so $\delta^{ij}_{LR}$ bound
is $\tan\beta$ independent, unlike all the other bounds.\\ 
In the GMI approximation, the amplitude has the following expression: 
\beqa
A^{ij}_{L1}=\frac{\alpha_{1}}{4\pi}
\frac{\Delta_{RL}^{ij}}{(m^{2}_{L}-m^{2}_{R})}\left(\frac{M_1}{m_{l_i}}\right)
\bigg(\frac{f_{3n}(a_R)}{m^2_{R}}-\frac{f_{3n}(a_L)}{m^2_{L}}\bigg).\nonumber
\eeqa
We note that the chirality flip is realized directly by the mass insertion so 
we can understand the order of the bound, compared to the $LL$ case,
$\delta_{ij}^{LR} \simeq(m_i/\tilde{m})\tan\beta\delta_{ij}^{LL}$, 
as confirmed numerically.\\ 
While MI and full computations give practically the same result both in $21$
and in $32$ sectors, the GMI approximation starts to work very well when $M_{1/2}\geq 300 GeV$.
This result can be better understood by bearing in mind the
conditions under which the GMI approximation can be applied. 
A necessary condition is that $M_{1,2}\geq m_W$ which is not satisfied
when $M_{1/2}\leq 300 GeV$.\\
In the $LL$ case we did not have this problem because of the chargino dominance.
We remind that $M_2\simeq 2M_1$ so we would expect the same behavior when 
$M_{1/2}\leq 150 GeV$ but this region is forbidden by the LSP Bino constraint.\\ 
In the $32$ sector, both MI and GMI approximations show a sizable deviation from 
the full computation in small $m_0$ regions because of a large $\delta^{33}_{RL}$.\\ 
A large $\delta^{33}_{RL}$ induces a mass split of order $\delta^{33}_{RL}$ 
in the third generation so, in the mass insertion language, 
we are neglecting next to leading terms of order
$\mathcal{O}((\delta^{{33}}_{RL})^2)$ not so suppressed in the examined case.
We note that the same argument holds for the $\delta^{32}_{LL}$ case. 

\begin{figure}[ht]
\begin{tabular}{cc}
\includegraphics[scale=0.35]{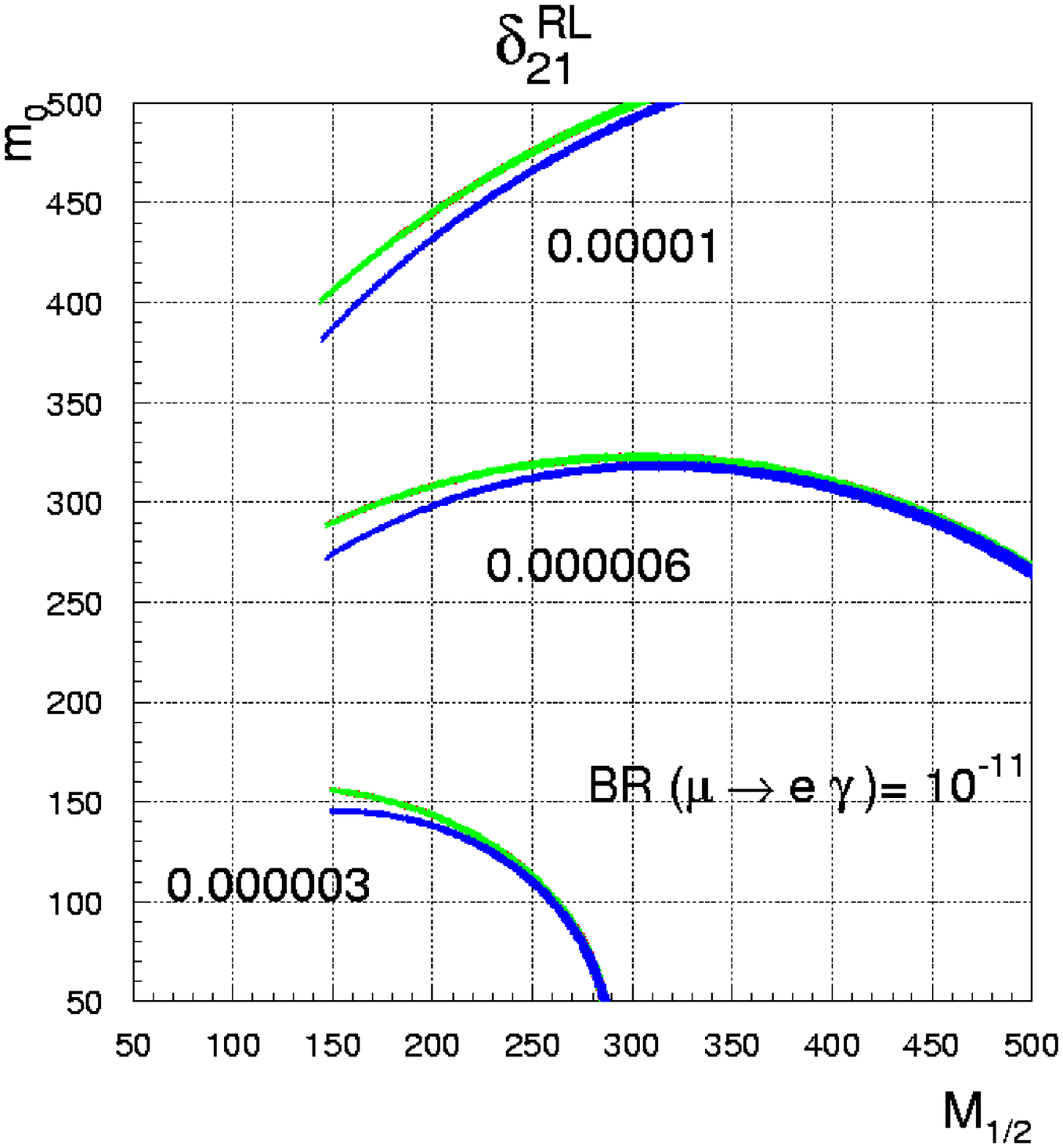} &
\includegraphics[scale=0.35]{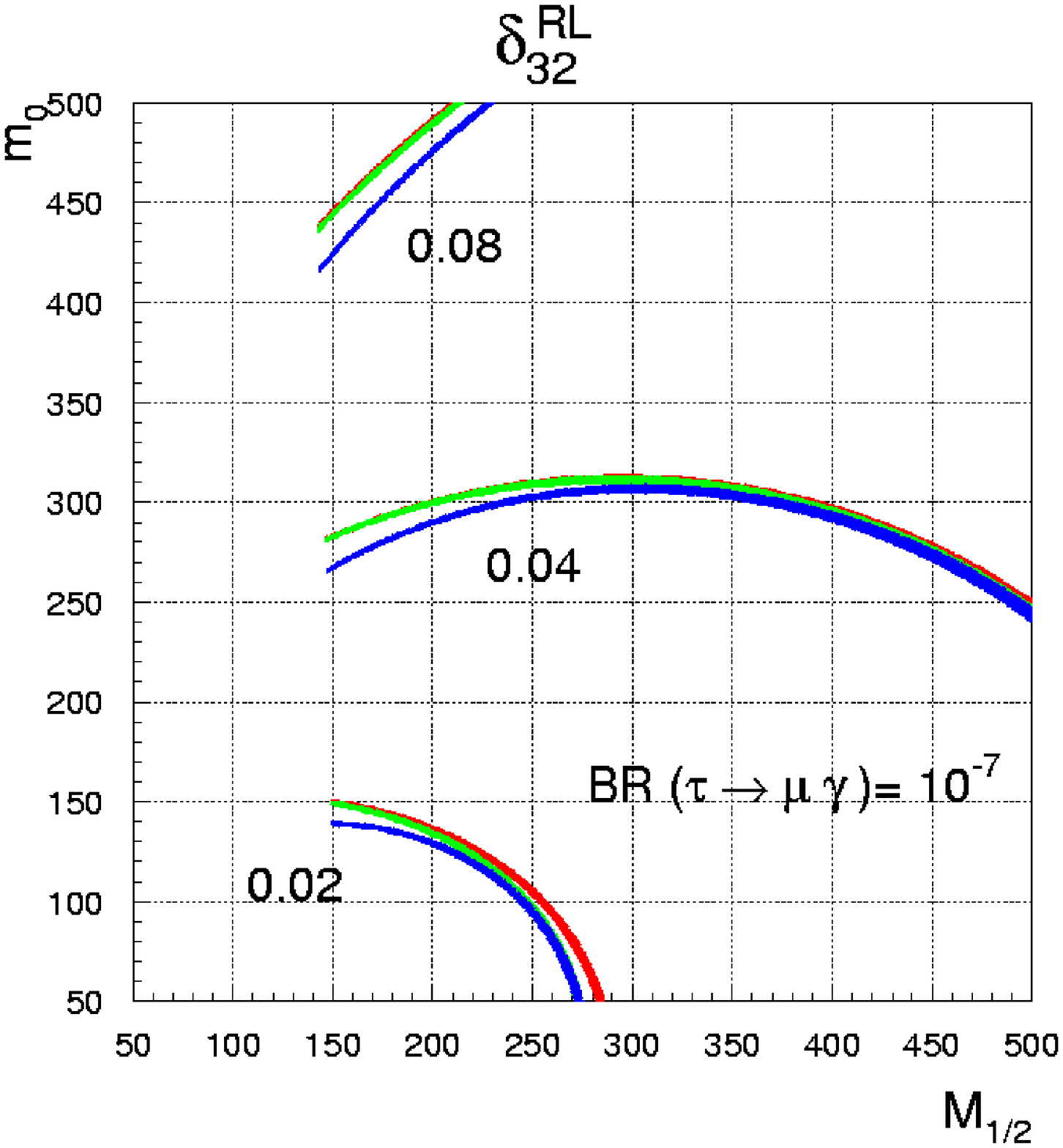}
\end{tabular}
\caption{Upper limits on $\delta^{ij}_{RL}$ from $l_i\rightarrow l_j\gamma$.
In the plots we set $\mu>0$, $\tan\beta=10$ and a=0. 
Red lines correspond to full computation,
green and blue lines to MI and GMI approximations respectively.}
\end{figure}
           
\subsection{\bf Bounds on $\delta_{RR}$}

The $\delta_{RR}$ sector requires a bit more of attention because of some 
cancellations occurring among the amplitudes in some regions of the
parameter space.\\ The bounds on $\delta_{RR}$ become very weak so, it is
interesting to check what is limit of applicability
of the mass insertion approximations \footnote{This typically occurs for RR type
MI as long as universality in the gaugino masses is maintained at the high scale.
Although in a completely generic situation without any universal boundary conditions, 
such cancellations can occur for LL type MI also \cite{stefanoyaguna1}.}.\\ 
The origin of this cancellations is the destructive interference between the dominant 
contributions coming from the $\tilde B$ 
(with chirality flip implemented through a FC mass insertion) 
and $\tilde B \tilde H^{0}$ exchange.
To better understand the nature of these cancellations 
let us derive, in the GMI approximation, the amplitude associated to $\delta_{RR}$:
\beqa
A^{ij}_R=\frac{\alpha_{1}}{4\pi}\Delta_{RR}^{ij}\,\,
&\bigg[&\frac{4f_{1n}(a_R)}{m^{4}_{R}}+
\mu M_{1}t_{\beta}\bigg(\frac{2f_{2n}(a_R,b_R)}{m^4_{R}(M_{1}^2\!-\!\mu^2)}+
\frac{1}{(m^{2}_{L}\!-\!m^{2}_{R})}\nonumber\\\nonumber\\
&\cdot&
\bigg(\frac{2f_{2n}(a_R)}{m^4_{R}}+\frac{1}{(m^{2}_{L}\!-\!m^{2}_{R})}
\bigg(\frac{f_{3n}(a_L)}{m^2_{L}}-\frac{f_{3n}(a_R)}{m^2_{R}}\bigg)\!\bigg)\!\bigg)\!\bigg].
\nonumber
\eeqa
\\
In the above equation, the first term is relative to the pure Bino amplitude with 
external chirality flip, the second one corresponds to the $\tilde B\tilde H^{0}$ mixing in 
the neutralino sector while the last term originates from the $\tilde{B}$ contribution with 
internal sfermion line chirality flip.\\  
It is easy to check numerically that the dominant contributions, 
proportional to $\tan\beta$, have opposite sign in all the parameter space.
To get a feeling of the reason of the above opposite sign, we note that  
the amplitude relative to $\tilde B$ (with chirality flip 
implemented through a FC mass insertion) is proportional to $N_{LL}N_{RR}$ with
$sign(N_{LL}N_{RR})=-1$ while the contribution arising from the $\tilde B\tilde H^{0}$ exchange
is proportional to $N_{LR}N_{RR}$ with $sign(N_{LR}N_{RR})=+1$ (see eqs.3).
Vice-versa, the same type of contributions in the $\delta_{LL}$ case have the same sign being
proportional to $N_{RR}N_{LL}$ with $sign(N_{RR}N_{LL})=-1$ and 
to $N_{RL}N_{LL}$ with $sign(N_{RL}N_{LL})=-1$, respectively. 
The above difference depends on the opposite sign between the hypercharge of 
SU(2) doublets and U(1) singlets.\\ 
If some cancellations occur among the leading contributions,
subleading effects, generally disregarded, could become important or even dominant.\\  
In this spirit, we retain the amplitude relative to a 
chirality flip realized in the external fermion line, 
neglected in \cite{MS} because it is not $\tan\beta$ enhanced.\\
Moreover, the above amplitude shows that while the dominant ($\tan\beta$ enhanced) 
contributions are proportional to the $\mu$ mass term
\footnote{In reality, this is true only if we neglect the $A$ term in
$\delta^{22}_{RL}\!=\!(A-\mu\tan\beta)m_{\mu}$ in the pure B amplitude.}
the pure $B$ amplitude with external chirality flip is $\mu$ independent .
In this way, the branching ratio is not invariant under the change of the $\mu$ sign
so, in general, one has to consider both cases.\\ 
In fig. 3 we show the upper limits on $\delta^{21}_{RR}$ from $\mu\rightarrow e\gamma$.
The limit on $\delta_{3j}^{RR}$ are simply obtained by 
$\delta_{3j}^{RR}/\delta_{21}^{RR}= 
(Br(\tau\rightarrow l_j\gamma)/Br(\mu\rightarrow e\gamma))^{1/2}$
thus, by now $\delta_{32}^{RR}$ and $\delta_{31}^{RR}$ are not constrained at all.\\ 
As we can see, we are not able to remove these cancellations, in fact, the effect of 
the $\tan\beta$ independent contribution is only a shift of the cancellation region 
(the same thing happens if we flip the $\mu$ sign).\\
\begin{figure}[ht]
\begin{center}
\includegraphics[scale=0.35]{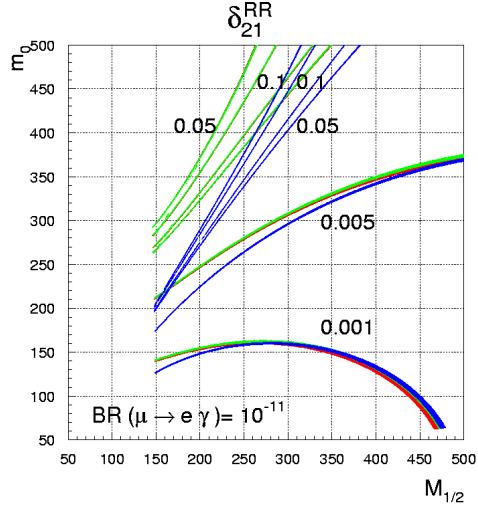} 
\caption{Upper limits on $\delta^{21}_{RR}$ from $\mu\rightarrow e\gamma$;
In the plots we set $\mu>0$, $\tan\beta=10$ and a=0 and red lines
correspond to the full computation,
green and blue lines to the MI and the GMI approximations, respectively.}
\end{center}
\end{figure}
We remark that, such cancellations occur in all the situations where the contribution of the 
$A$ term to $\delta^{22}_{RL} = (A-\mu\tan\beta)m_{\mu}$ is negligible.
There are well known model dependent upper bounds on the $A$ parameter to avoid
color and e.m. charge breaking, in particular $|A|/m_R\leq 3$ in mSUGRA.
Moreover, mSUGRA requires a large $\mu$ term to fulfill the electroweak symmetry breaking,
thus, we cannot invert the relative sign between the two amplitudes.
In this spirit, we neglected the $A$ term in $A^{ij}_R$.
It is noteworthy that the MI approximation works very well reproducing the same 
cancellation regions as the full computation. 
In the GMI case, we have a net shift of this region but the general structure is maintained.\\ 
In conclusion, $\mu\rightarrow e\gamma$ doesn't allow to put a bound in the $RR$ sector,
so, we take into account other LFV processes as 
$\mu\rightarrow  eee$ and $\mu\rightarrow e\ in \ Nuclei$.\\
As fig. 5 shows, we find that the last process suffers from 
a bigger cancellation problem than $\mu\rightarrow  e\gamma$ (in the $RR$ sector) 
while $\mu\rightarrow eee$ does not.\\
This result requires some explanations. As we have seen in sec.5, the dominant 
contribution to $Br(\mu\rightarrow e\ in \ Nuclei)$ and $Br(\mu\rightarrow eee)$
arises from the dipole operator (that is $\tan\beta$ enhanced) but, if the dipole
amplitude is strongly suppressed by some cancellations,  
$Br(\mu\rightarrow e\gamma)$ goes to zero while $Br(\mu\rightarrow e\ in \ Nuclei)$ and 
$Br(\mu\rightarrow eee)$ are dominated by non-dipole contributions. 
So, in principle, $\mu\rightarrow e\ in \ Nuclei$
and $\mu\rightarrow eee$ could be able to bound $\delta_{RR}$.
As we can see in fig.5, this is the case of $\mu\rightarrow eee$ that gives a bound for 
$\delta_{RR}\leq 0.4$ that is, correctly, $\tan\beta$ independent.\\  
On the other hand, we find that $Br(\mu\rightarrow e\ in \ Nuclei)$ has additional cancellations 
between dipole and not-dipole amplitudes.
As a final effect, we have that $\mu\rightarrow e\ in \ Nuclei$ suffers from 
a similar cancellation problem as $\mu\rightarrow e\gamma$ but,
as can be expected, in a different region of the parameter space.\\ 
Because of this strong cancellations, $\mu \rightarrow e \gamma$ and
$\mu\rightarrow e\ in \ Nuclei$ prevent us from getting a bound in
the $RR$ sector both at the present and even in the future when their
experimental sensitivity will be improved.\\
However, an interesting feature is that $\mu\rightarrow e\gamma$ and 
$\mu\rightarrow e\ in \ Nuclei$ amplitudes have cancellations in different regions, so, if 
we combine the two processes, we obtain a more stringent bound ($\delta_{21}^{RR}\leq$  0.2)
than the one coming from $\mu\rightarrow eee$ ($\delta_{21}^{RR}\leq$  0.4).  
It is noteworthy that the study of combined processes allows to extract additional
informations respect to each separate case.\\ 

\begin{figure}[ht]
\begin{tabular}{cc}
\includegraphics[scale=0.35]{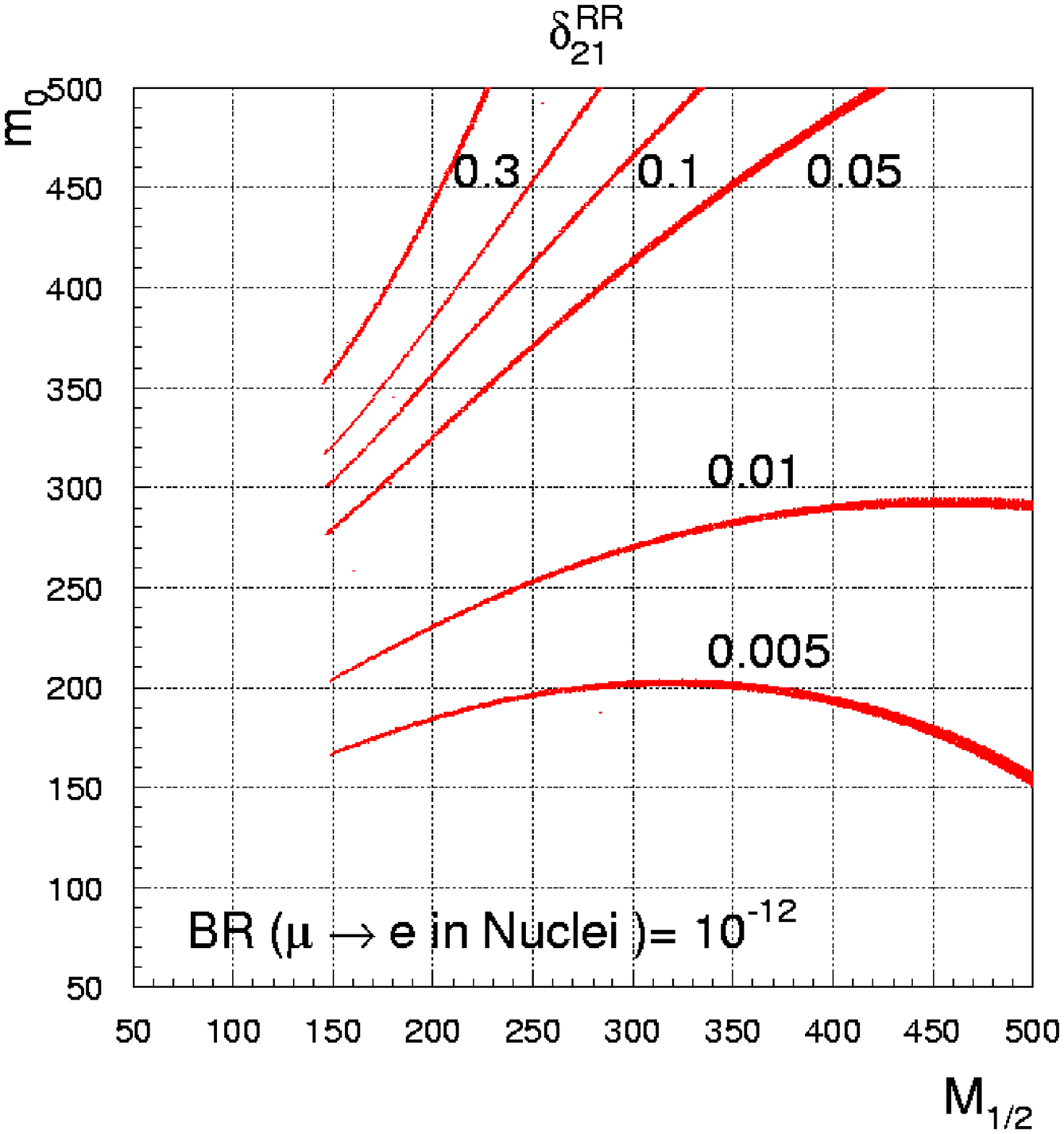} &
\includegraphics[scale=0.35]{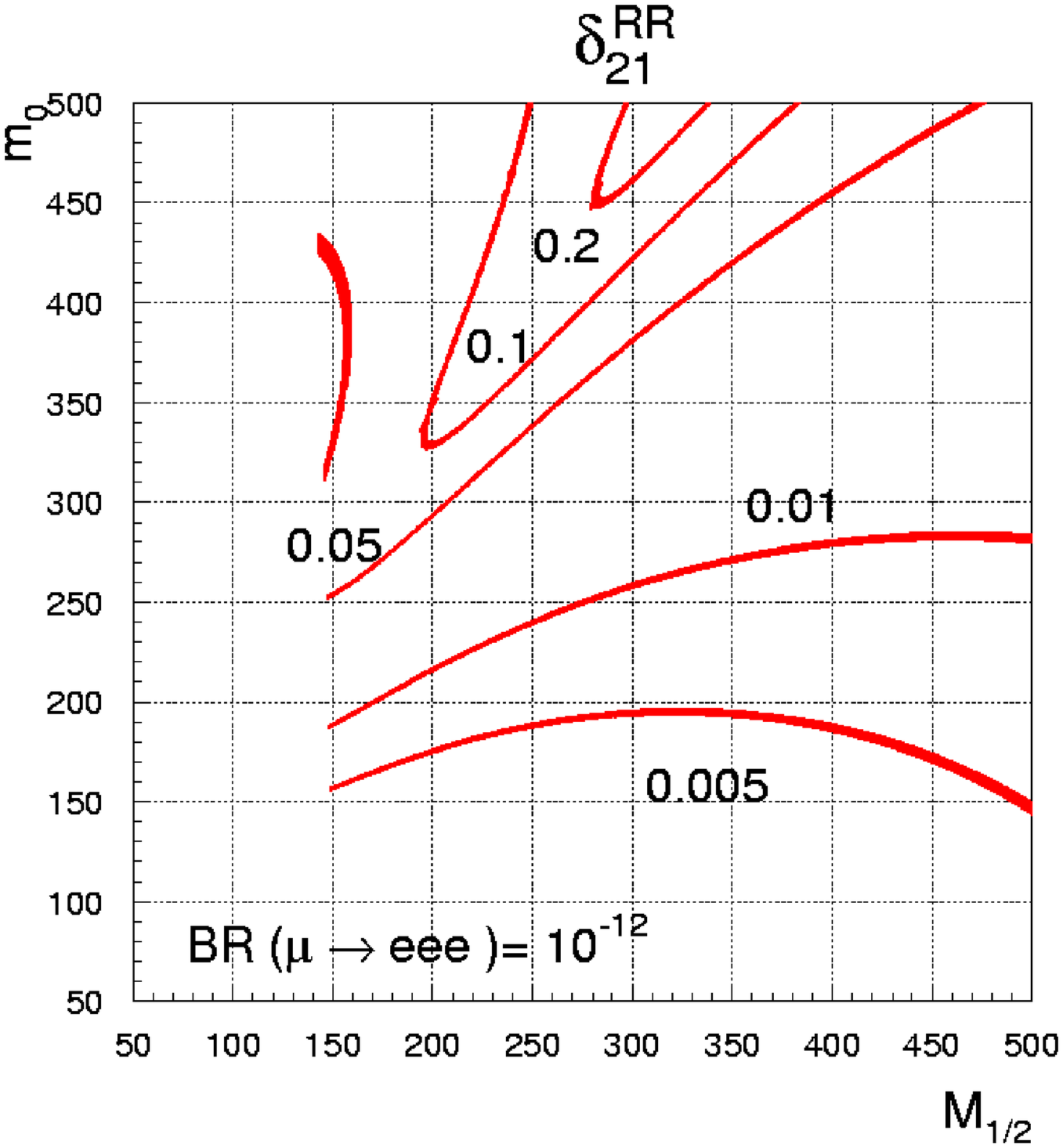}
\end{tabular}
\caption{Upper limits on $\delta^{21}_{RR}$ from $\mu\rightarrow e \ in \ Ti$ (a) and 
$\mu\rightarrow eee$ (b). In the plots we set $\mu> 0$, $\tan\beta=10$ and a=0.}
\end{figure}
\begin{figure}[ht]
\begin{tabular}{cc}
\includegraphics[scale=0.35]{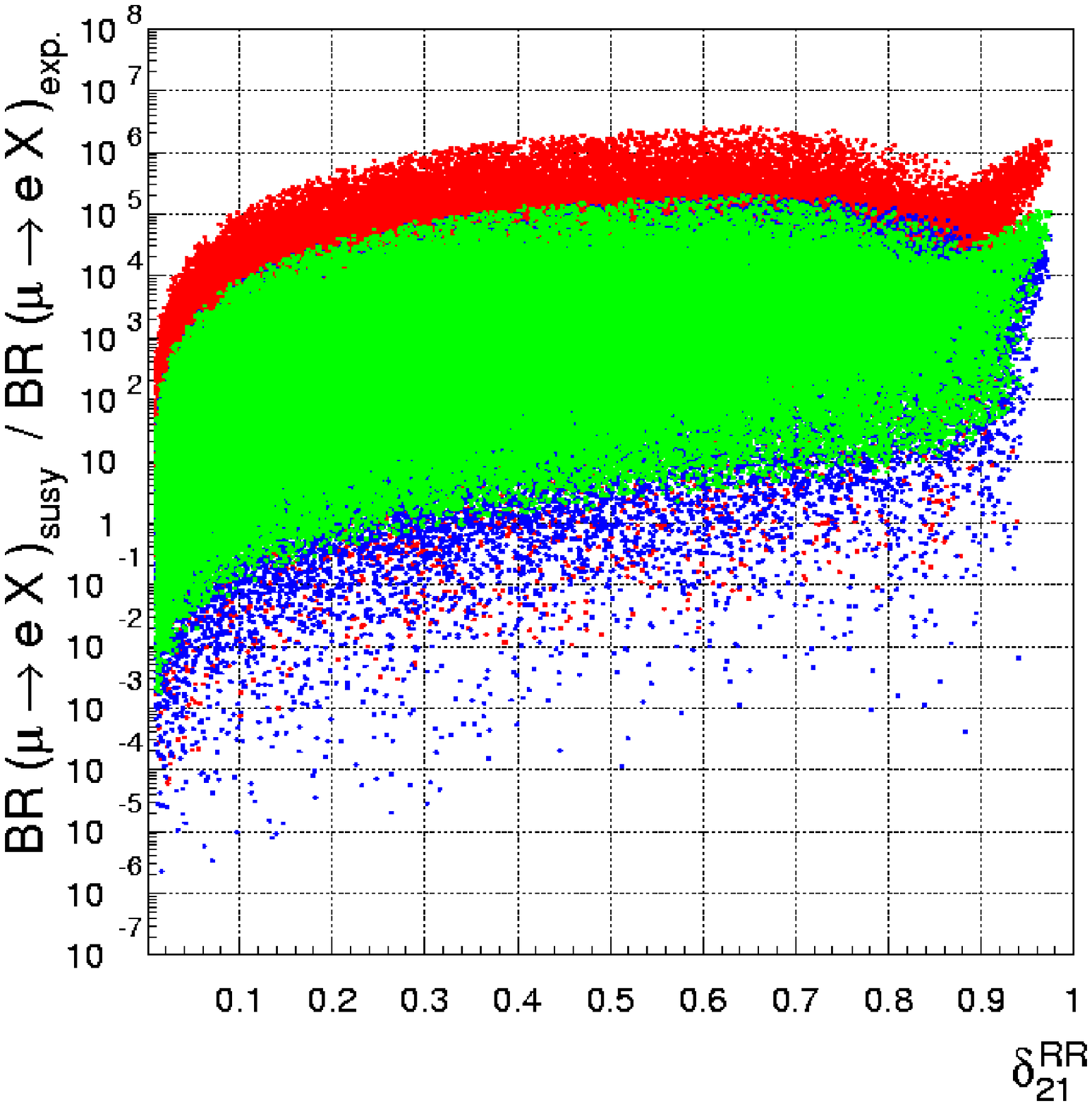} &
\includegraphics[scale=0.35]{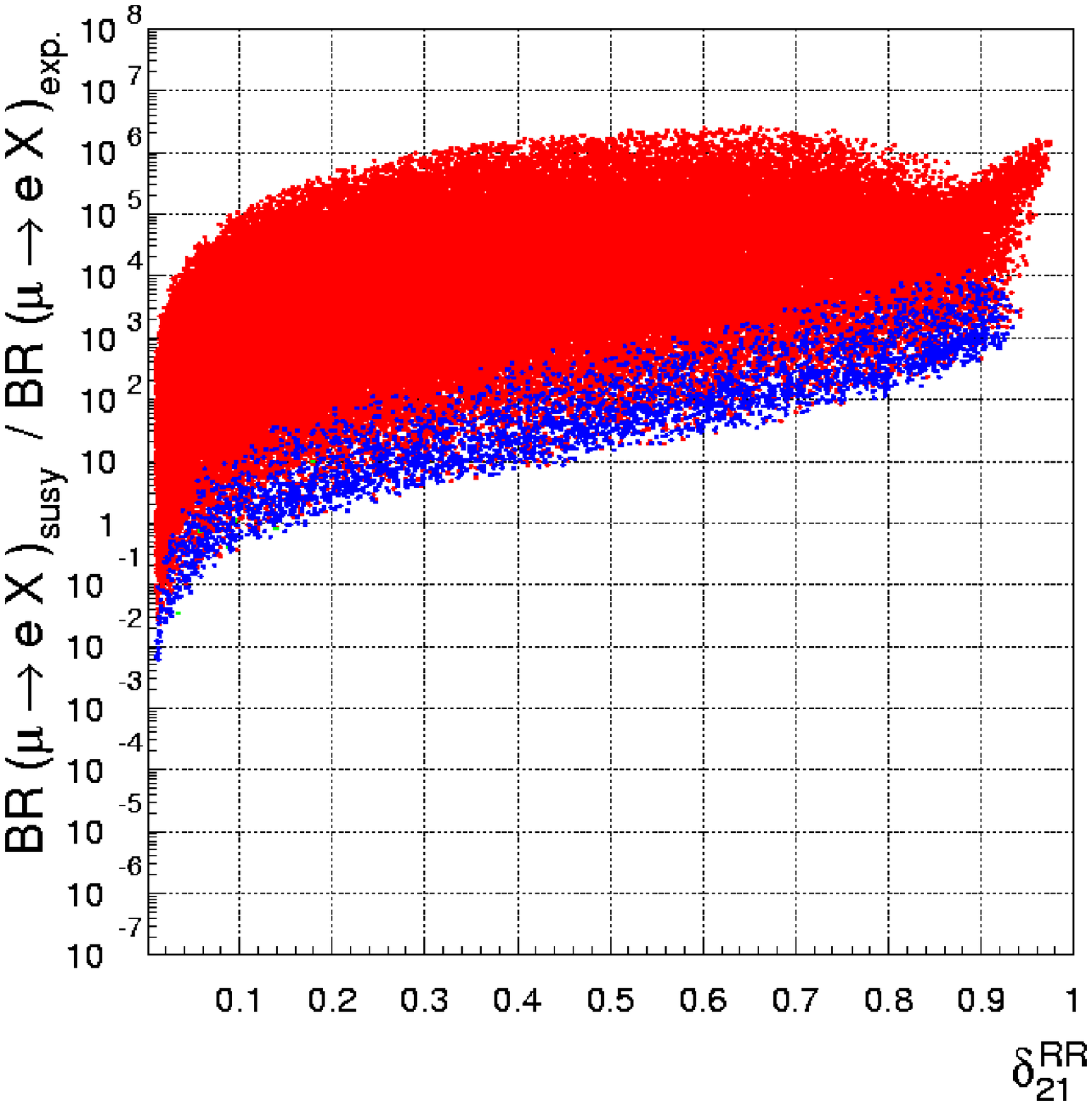}
\end{tabular}
\caption{Branching ratios of $\mu\rightarrow e$ transitions normalized to the actual 
experimental upper bounds vs $\delta^{21}_{RR}$. 
Red dots correspond to X = $\gamma$, blue dots to X = ``in Nuclei'' 
and green dots to X = ee. In the scatter plots we have taken $\mu> 0$,  $3<\tan\beta<40$
50\ GeV$\leq m_{0}\leq$500\ GeV, 50\ GeV $\leq M_{1/2}\leq$ 500\ GeV and $-3<a<3$.
On the right, for each point of the parameter space,
it is shown only the biggest normalized Br value
among $\mu\rightarrow e \gamma$, $\mu\rightarrow eee$ and $\mu\rightarrow e \ in \ Ti$.}
\end{figure}

\subsection{\bf Bounds on the double mass insertion $\delta^{32}$ $\delta^{31}$ 
from $\mu\rightarrow e \gamma$}

As we have seen in the previous section, the bounds on $\tau\rightarrow \mu \gamma$ and 
$\tau\rightarrow \e \gamma$ are very loose. This is due to a worse experimental resolution 
on the above processes compared to the $\mu\rightarrow \e \gamma$ one.
However, we can extract additional information in the $32$ and $31$ sectors
applying to $\mu\rightarrow e \gamma$.\\ The point is that we are able to put bounds on
the product of two mass insertions, namely $\delta_{32}\delta_{31}$.
In general, we can pass from the second to the first generation or
through the $\delta_{21}$ or through the $\delta_{23}\delta_{31}$ insertions.\\
Now, to constraint the MIs, we proceed exactly as for a single MI.
We put to zero all off diagonal FV entries in the slepton mass matrix except 
for the two MIs we are interested to bound. 
At this point, we give the analytical expressions for all the $\delta_{23}\delta_{31}$ type insertions 
in the GMI approach.\\ For $\delta^{RL}_{23}\delta^{LL}_{31}$ we get the following amplitude:
\beqa
A^{21}_{L1}\!=\!\frac{\alpha_{1}}{4\pi}\frac{M_{1}}{m_{\mu}}
\frac{\Delta_{RL}^{23}\Delta_{LL}^{31}}{(m^{2}_{L}\!-\!m^{2}_{R})}\,\,
\bigg[\frac{2f_{2n}(a_L)}{m^4_{L}}+\frac{1}{(m^{2}_{R}\!-\!m^{2}_{L})}
\bigg(\frac{f_{3n}(a_R)}{m^2_{R}}\!-\!\frac{f_{3n}(a_L)}{m^2_{L}}\bigg)\bigg].\nonumber
\eeqa
The amplitude $A^{21}_{R1}$, relative to $\delta^{LR}_{23}\delta^{RR}_{31}$,
is obtained by $A^{21}_{R1}=A^{21}_{L1}(L\!\leftrightarrow\!R)$.
In fig. 6 we show the bounds on $\delta^{RL}_{23}\delta^{LL}_{31}$ and on
$\delta^{LR}_{23}\delta^{RR}_{31}$.\\ As we can see, they exhibit different behaviors,
especially for $m_0$ smaller than $M_{1/2}$ due to the $m_R$ and $m_L$ mass difference
(in fact, while in the first case we have two left handed and one right handed sfermions
running in the loop, in the second case we have the opposite situation).\\ 
So, while $\delta^{ij}_{RL}$ and $\delta^{ij}_{LR}$ are indistinguishable,
it is not so for $\delta^{LR}_{23}\delta^{RR}_{31}$ and $\delta^{RL}_{23}\delta^{LL}_{31}$.\\
The amplitude associated to $\delta^{LL}_{23}\delta^{LL}_{31}$, 
namely $A^{21}_{L2}=(A^{21}_{L2})_{SU(2)}+(A^{21}_{L2})_{U(1)}$, reads: 
\beqa
\left(A^{21}_{L2}\right)_{SU(2)}\!=\!\frac{\alpha_{2}}{4\pi}\Delta_{LL}^{23}\Delta_{LL}^{31}
&\bigg[&\frac{I_{1n}(a_L)\!+\!I_{1c}(a_L)}{m^6_{L}}\!+\!
\frac{\mu M_{2}t_{\beta}}{(M_{2}^2\!-\!\mu^2)}\nonumber\\\nonumber\\
&\cdot&\frac{I_{2n}(a_L,b_L)\!+\!I_{2c}(a_L,b_L)}{m^6_{L}}\bigg]\nonumber
\eeqa
\beqa
&\left(A^{21}_{L2}\right)_{U(1)}&=\frac{\alpha_{1}}{4\pi}\Delta_{LL}^{23}\Delta_{LL}^{31}
\bigg[\frac{I_{1n}(a_L)}{m^6_{L}}\!+\!
\mu M_{1}t_{\beta}\bigg(\frac{-I_{2n}(a_L,b_L)}{m^6_{L}(M_{1}^2\!-\!\mu^2)}\!+\!
\frac{1}{(m^{2}_{R}\!-\!m^{2}_{L})}\nonumber\\\nonumber\\
&\cdot\,\bigg(&\!\!\!\!\!\!\!\!\!\!\frac{2I_{2n}(a_L)}{m^6_{L}}
+\frac{2f_{2n}(a_L)}{m^{4}_{L}(m^{2}_{R}-m^{2}_{L})}+\frac{1}{(m^{2}_{R}-m^{2}_{L})^2}
\bigg(\frac{f_{3n}(a_R)}{m^2_{R}}-\frac{f_{3n}(a_L)}{m^2_{L}}\bigg)\!\bigg)\!\bigg)\!\bigg].\nonumber
\eeqa \\
The contribution arising from a $\delta_{LL}^{23}\delta_{RR}^{31}$ 
type insertion reads:
\beqa
A^{21}_{L3}=&-&2\,\frac{\alpha_{1}}{4\pi}\,\mu M_{1}t_{\beta}
\frac{\Delta_{LL}^{23}\Delta_{RR}^{31}}{(m^{2}_{L}\!-\!m^{2}_{R})^2}\,\,
\bigg[\frac{f_{2n}(a_L)}{m^4_{L}}+\frac{f_{2n}(a_R)}{m^4_{R}}\nonumber\\\nonumber\\
&+&
\frac{1}{(m^{2}_{R}\!-\!m^{2}_{L})}
\bigg(\frac{f_{3n}(a_R)}{m^2_{R}}\!-\!\frac{f_{3n}(a_L)}{m^2_{L}}\bigg)\bigg].\nonumber
\eeqa
In the fig. 7 we show the bounds for $\delta^{LL}_{23}\delta^{LL}_{31}$ and for
$\delta^{LL}_{23}\delta^{RR}_{31}$ (equal to the bounds on $\delta^{RR}_{23}\delta^{LL}_{31}$).
It is to note that $\delta^{LL}_{23}\delta^{RR}_{31}$ is strongly constrained   
because, the associate amplitude, is $m_{\tau}/m_{\mu}$ enhanced respect to the usual 
Bino-like mediated processes (being the chirality flip implemented in
the internal sfermion line through $\delta^{LR}_{33}$
and not by $\delta^{LR}_{22}$, as usual). 
The amplitude $A^{21}_{R3}$, relative to $\delta^{RR}_{23}\delta^{LL}_{31}$,
is $A^{21}_{R3}=A^{21}_{L3}(L\leftrightarrow R)$.\\
Finally we derive the expression for the amplitude associated to $\delta^{RR}_{23}\delta^{RR}_{31}$:
\beqa
\left(A^{21}_{R2}\right)&=&\frac{\alpha_{1}}{4\pi}
\Delta_{RR}^{23}\Delta_{RR}^{31}\,\,
\bigg[\frac{4I_{1n}(a_R)}{m^6_{R}}+
\mu M_{1}t_{\beta}
\bigg(\frac{2I_{2n}(a_R,b_R)}{m^6_{R}(M_{1}^2\!-\!\mu^2)}+
\frac{1}{(m^{2}_{R}\!-\!m^{2}_{L})}\nonumber\\\nonumber\\
&\cdot\,\bigg(&\!\frac{-2I_{2n}(a_R)}{m^6_{R}}\!+\!
\frac{2f_{2n}(a_R)}{m^{4}_{R}(m^{2}_{R}\!-\!m^{2}_{L})}\!+\!\frac{1}{(m^{2}_{R}\!-\!m^{2}_{L})^2}
\bigg(\frac{f_{3n}(a_R)}{m^2_{R}}\!-\!\frac{f_{3n}(a_L)}{m^2_{L}}\bigg)\!\bigg)\!\bigg)\!\bigg]\nonumber.
\eeqa
We note that, in general, $\delta_{23} \delta_{31}$ has a bound comparable to $\delta_{21}$, 
being $\delta_{21}\simeq \delta_{23} \delta_{31}$.
The only exception is for $\delta^{RR}_{23}\delta^{LL}_{31}$ ($\delta^{LL}_{23}\delta^{RR}_{31}$)
as discussed above.
\begin{figure}[ht]
\begin{tabular}{cc}
\includegraphics[scale=0.35]{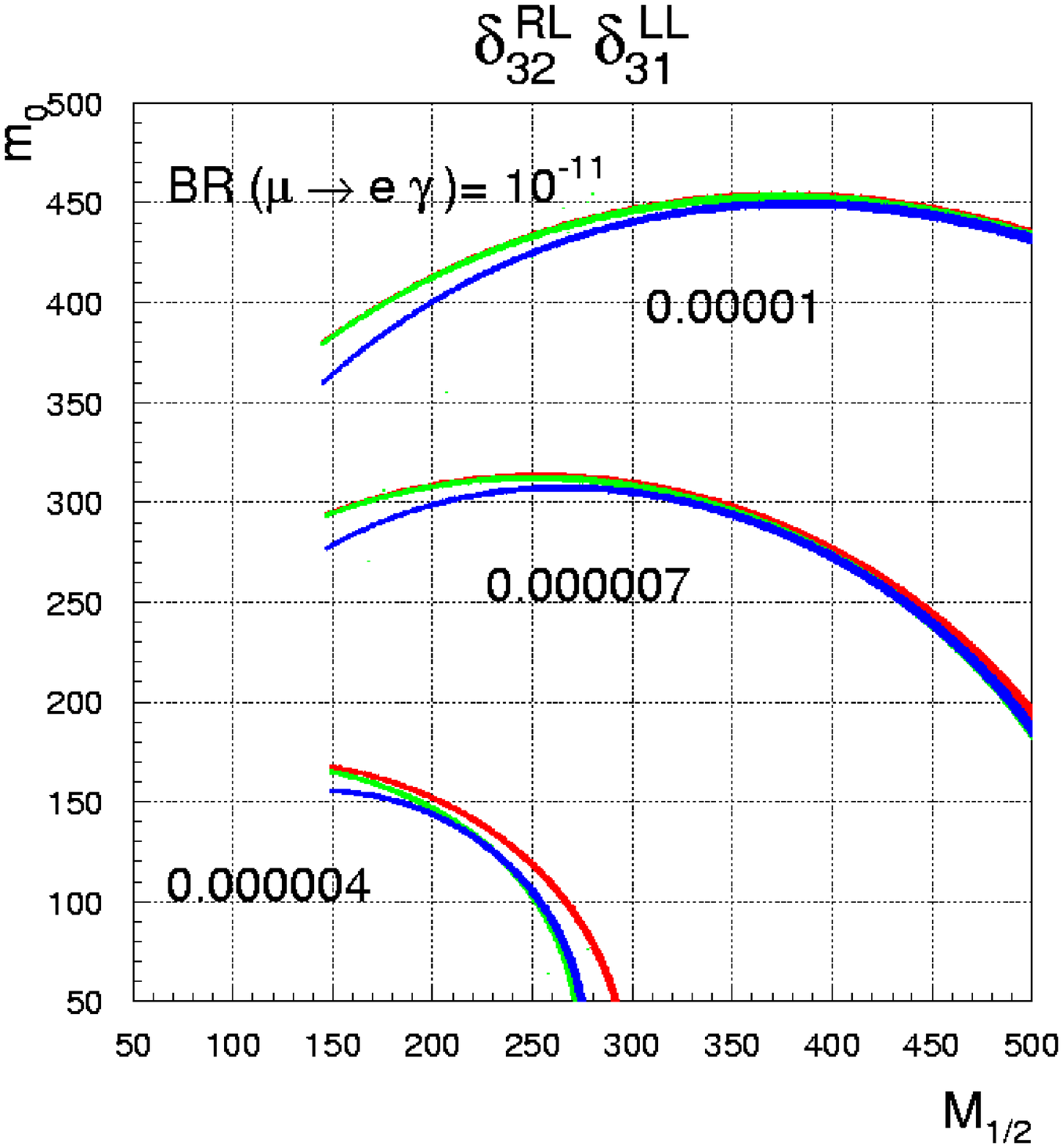} &
\includegraphics[scale=0.35]{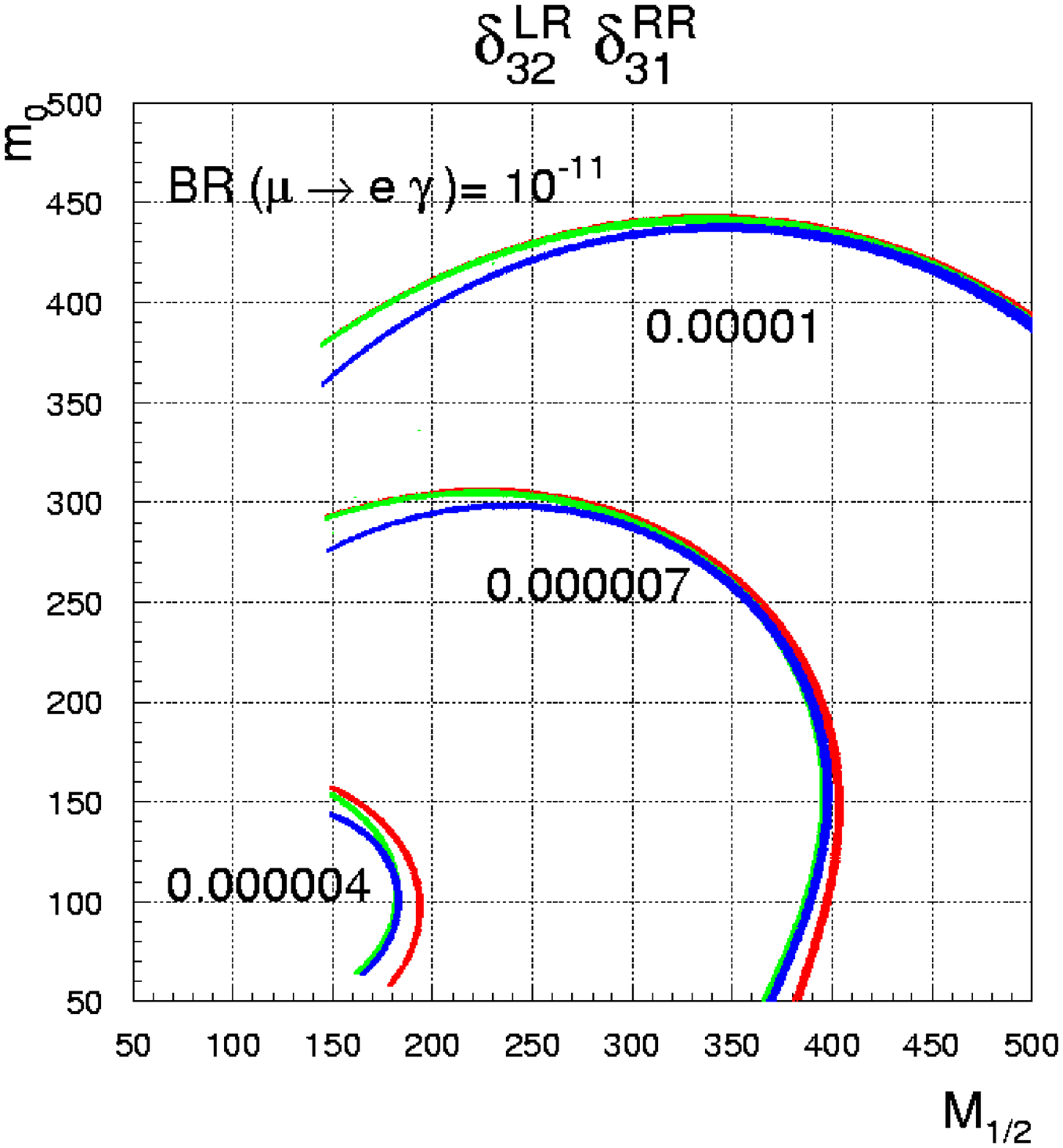}
\end{tabular}
\caption{Upper limits on $\delta^{32}_{RL}$ $\delta^{31}_{LL}$ and $\delta^{32}_{LR}$ $\delta^{31}_{RR}$ 
from $\mu\rightarrow e \gamma$. We have set $\mu> 0$, $\tan\beta=10$ and a=0. 
Red lines correspond to full computation, green and blue lines to MI and GMI approximations respectively.}
\end{figure}

\begin{figure}[ht]
\begin{tabular}{cc}
\includegraphics[scale=0.35]{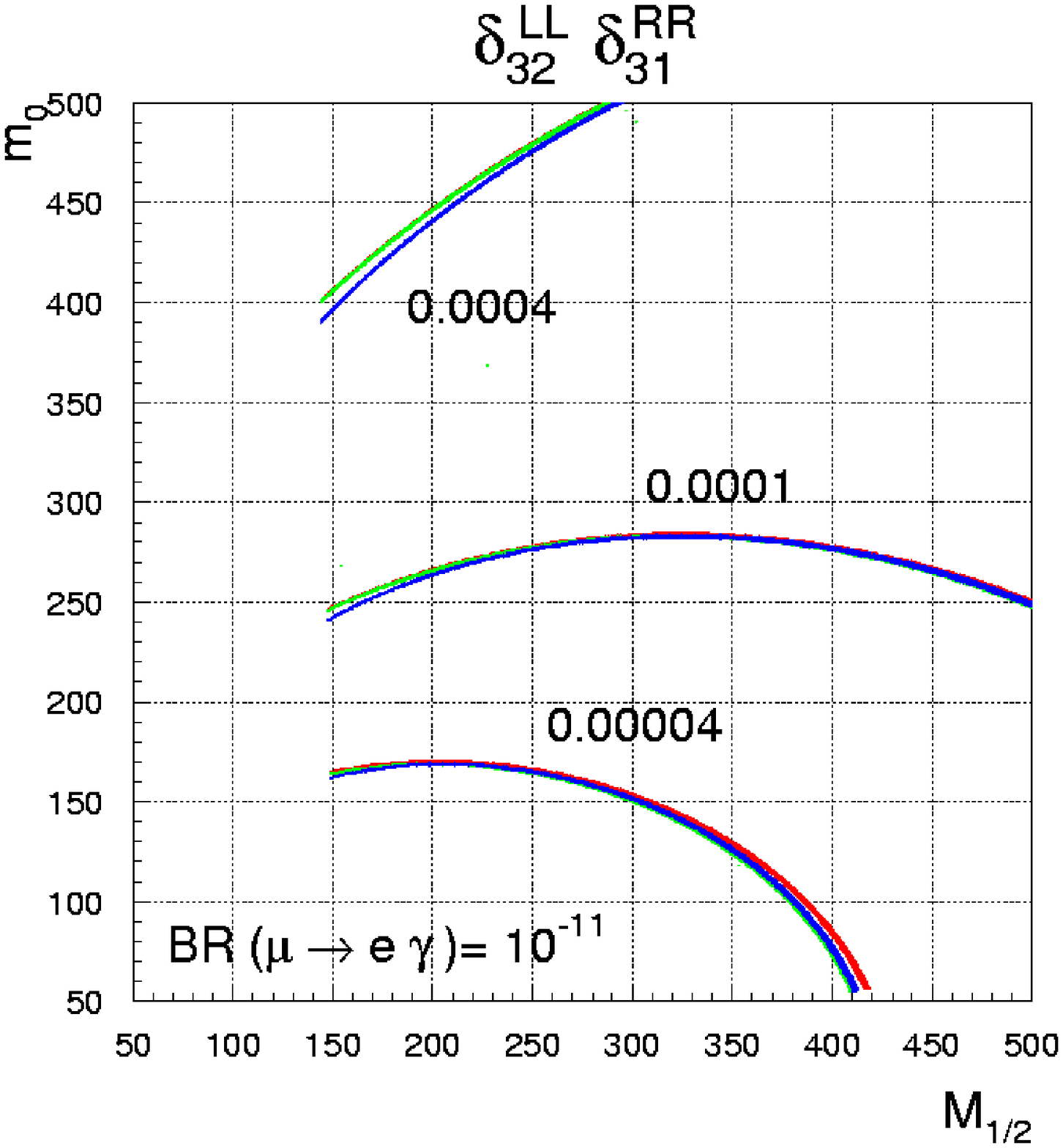} &
\includegraphics[scale=0.35]{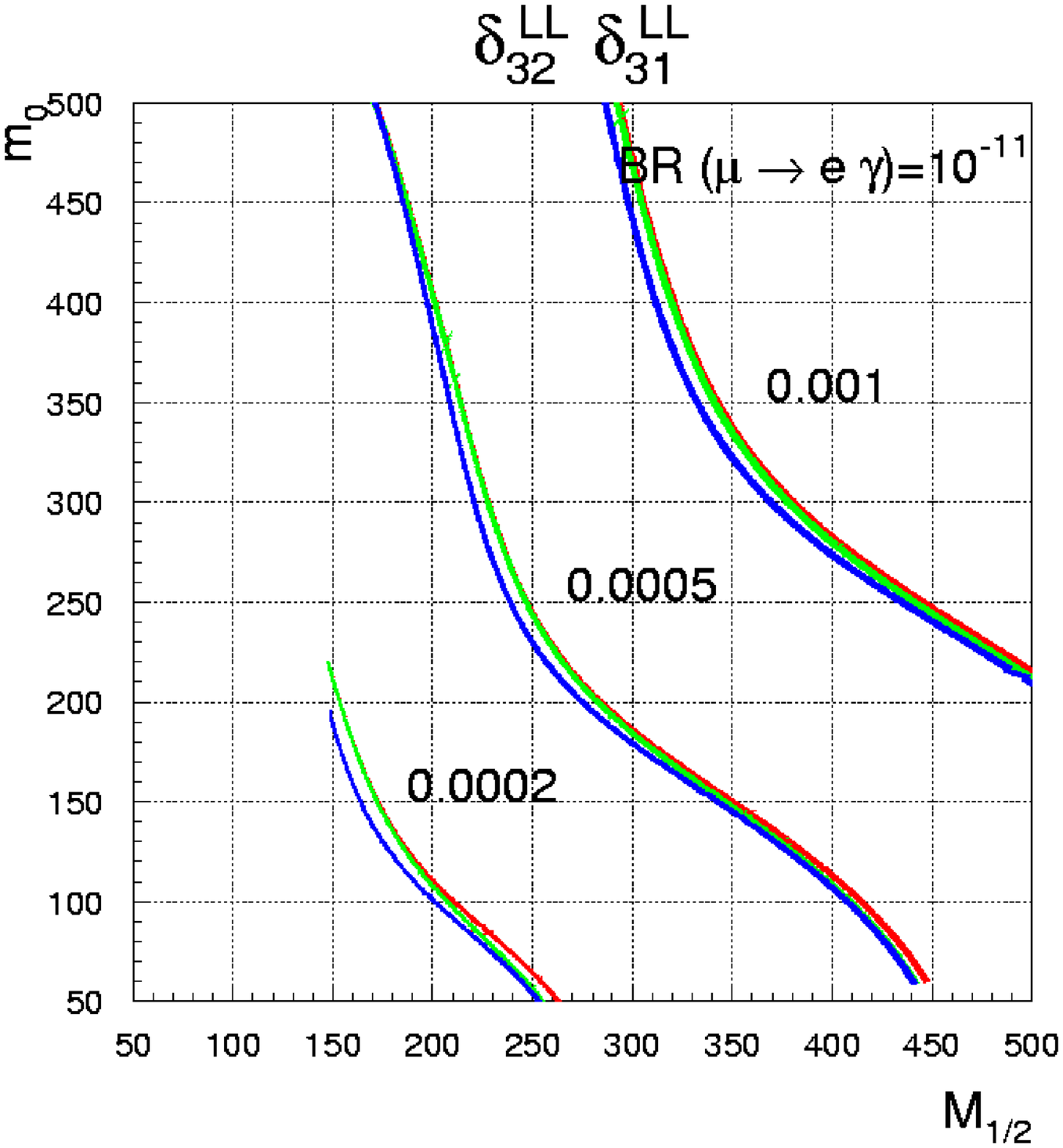}
\end{tabular}
\caption{Upper limits on $\delta^{32}_{LL}$ $\delta^{31}_{RR}$ and $\delta^{32}_{LL}$ $\delta^{31}_{LL}$ 
from $\mu\rightarrow e \gamma$. We have set $\mu> 0$, $\tan\beta=10$ and a=0. Red lines correspond to the 
full computation, green and blue lines to the MI and to the GMI approximations respectively.}
\end{figure}

\begin{figure}[ht]
\begin{tabular}{cc}
\includegraphics[scale=0.35]{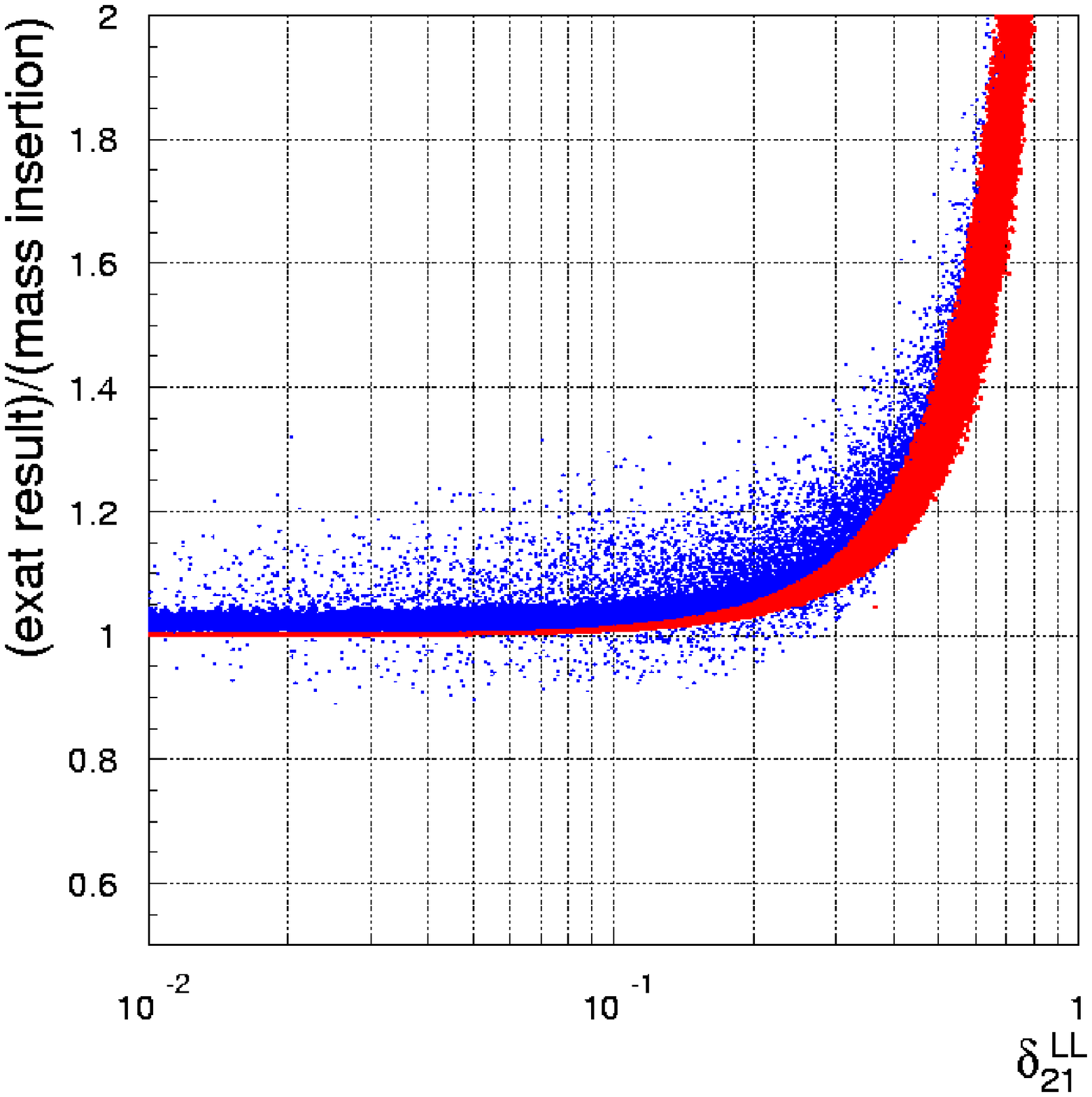} &
\includegraphics[scale=0.35]{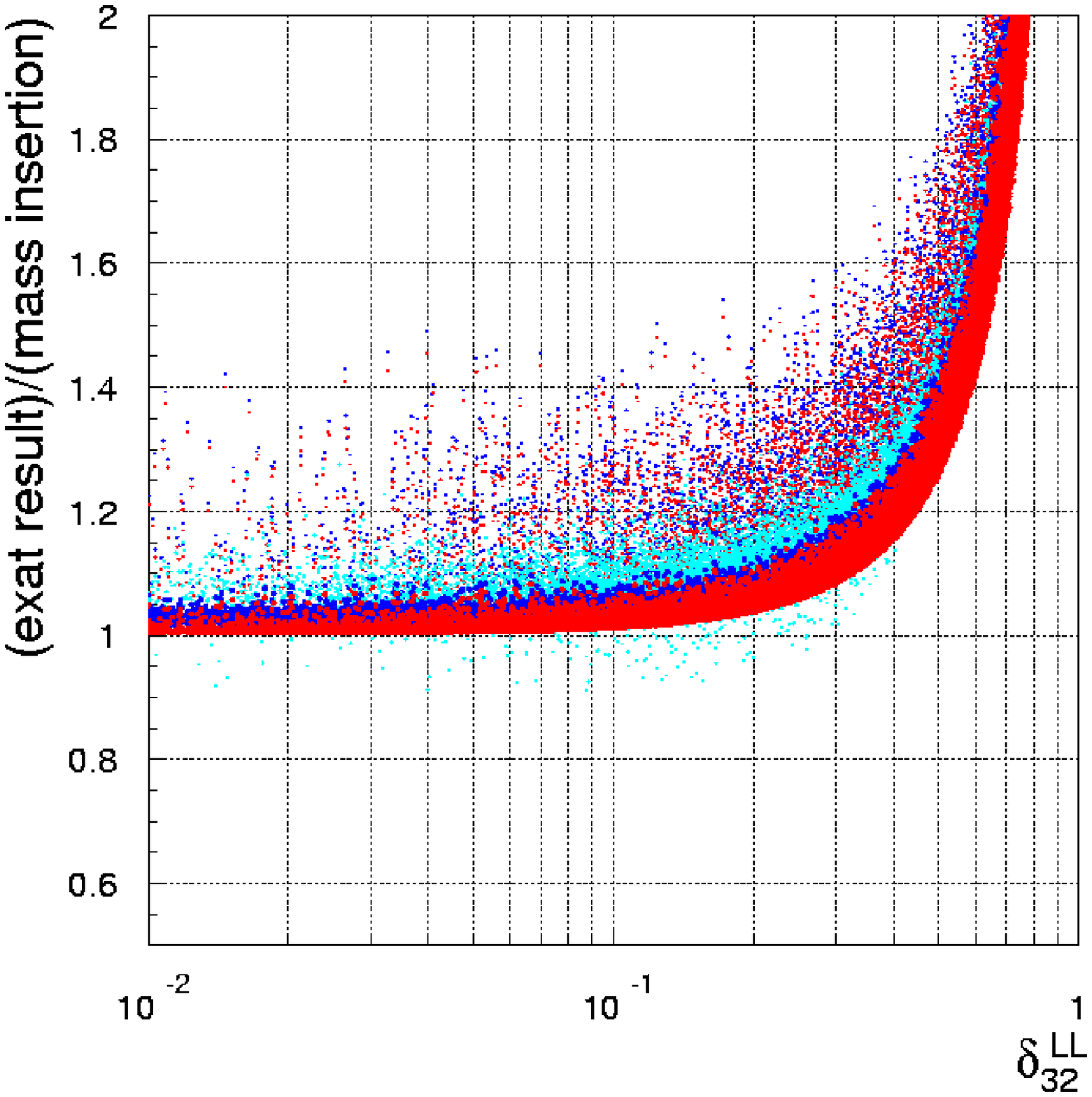}
\end{tabular}
\caption{ Ratios between exact and MI approximation (red dots) 
and between exact and GMI approximation (blue dots) vs $\delta^{ij}_{LL}$.
On the right, tiny light blue dots are relative 
to the chargino-neutralino approximations while tiny blue and red dots are due to large 
FC terms effects.}
\end{figure}

\section{\bf Mass eigenstate vs mass insertion: a numerical analysis}
 
At this point, our purpose is to make a quantitative estimate of the goodness of the MI and GMI 
approximations.\\
Even if we have already seen that the above approximations give us the same bounds  
as the full computation, we were not able, in the previous analyses, to quantify and 
to distinguish correctly the approximations induced by the chargino (neutralino) branch and   
by the slepton branch. In the slepton case, we can distinguish between two sources of 
approximations, namely, FC and FV terms.\\
In fig. 8 we show the ratios between full/MI computations (red dots)
and full/GMI computations (blue dots) vs. FV mass insertions.\\
As we can see in the $21$ sector, MI approximates correctly 
(at $(5\!-\!10)\%$ level) 
the full computation until large FV insertions ($\delta_{21}\simeq 0.2$),
while the two computations are practically indistinguishable for $\delta_{21}\leq 0.1$.
The induced approximations are easily understood if we remind that the approach we followed 
to put the bounds on the $\delta^{ij}$ insertions was to consider only one mass insertion contributing 
at a time for the $Br(l_i\rightarrow l_j\gamma)$.
If we stop in the first term we neglect terms of order $\delta^3_{ij}$ 
in the amplitudes inducing, naively, an approximation on the branching ratios
$(Br(\delta_{ij}+\delta^3_{ij})-Br(\delta_{ij}))/
Br(\delta_{ij}+\delta^3_{ij})\sim 2\delta^2_{ij}$,
as it is well reproduced numerically. The FC insertion $\delta^{LR}_{22}$ does not produce 
any sizable effects, in fact, in the worse case (for large $\mu$ and $\tan\beta$ and 
moderate slepton masses), we have $\delta^{LR}_{22}\leq 0.1$.
The last argument is not true in the $32$ sector where, being 
$\delta^{LR}_{33}/\delta^{LR}_{22}\simeq m_{\tau}/m_{\mu}$, we can have a not perturbative 
FC insertion. This is clear in fig. 8 where, the MI and the GMI approaches 
underline a $(40-50)\%$ deviations respect to the exact calculation 
(tiny dots refer to the $\tan\beta\geq 30$ and $M_{1/2}\geq 300$, 
region where we have sizable deviations due to the slepton FC terms only).\\   
Now, we want to discuss the approximations brought from the chargino and the neutralino sectors 
in the GMI approach (the following discussion is obviously flavour independent).\\
As discussed in section 3, it is allowed to use this method when the elements outside 
from the diagonal (proportional to $m_W$) are much smaller than those
diagonals ($\mu$ and $M_{2,1}$).\\
On the left of fig. 8, tiny blue dots show a $(20-30)\%$ approximation (it happens 
in $M_{1/2} \leq 300$ and $\tan\beta \leq 10$) 
while the much larger ones refer to the $M_{1/2} \geq 300$ and $\tan\beta \geq 10$ 
region where the GMI conditions are fulfilled.
In the $32$ case light blue dots correspond to deviations induced by the GMI approximation
in fact they are related to a range of parameters ($m_0\geq 300$ and $\tan\beta \leq 5$)
where $\delta^{LR}_{33}\leq 0.1$.\\
To summarize the results found, we can say that MI approximation produce 
the same features as the exact calculation even if strong cancellations occur.\\  
The approach works better in the $21$ sector than in the $32$ one because, while
in the first case we always have perturbative FC terms, in the second case it is not so
and we can reach sizable deviations (up to a $(40-50)\%$ level) from the full computation.\\
Moreover, we have a $10\%$ approximation until FV terms of order 0.2.\\
The GMI approximation works very well, as the MI approximation, up to gaugino masses
heavier than $150 GeV$ and it produces the cancellations
in a shifted region respect to the exact case.\\
In conclusion we can say that, except for fine-tuned cases,
the last approach is very satisfactory. 

\section{Conclusions}

In this work, in the first stage, we studied the constraints on
flavour violating terms in low energy SUSY coming from $l_i\rightarrow l_j \gamma$.\\ 
We have carried out the analysis both in the mass eigenstate and in
the mass insertion approximations clarifying the limit of
applicability of these approximations.\\ 
In particular, we focused on the RR sector where strong cancellations 
make the sector unconstrained.\\ 
We showed that these cancellations prevent us from getting a bound in
the $RR$ sector both at the present and even in the future when the
experimental sensitivity on $Br(l_i \rightarrow l_j \gamma)$ will be improved.\\
Finally we took into consideration the bounds on the various double mass insertions:   
$\delta_{23}^{LL} \delta_{31}^{LL}$, $\delta_{23}^{RR} \delta_{31}^{RR}$,
$\delta_{23}^{LR} \delta_{31}^{RR}$, $\delta_{23}^{RL} \delta_{31}^{LL}$ and
$\delta_{23}^{RR} \delta_{31}^{LL}$. 
It is clear that the bounds are approximatively the same as in $\delta_{21}$ being
$\delta_{21}\simeq \delta_{23} \delta_{31}$
except for the last one suppressed by a $m_{\mu}/m_{\tau}$ factor.
So, in spite of very weak bounds on $\delta_{32}$ and on $\delta_{31}$
coming from $Br(\tau\rightarrow \mu \gamma)$ and $Br(\tau\rightarrow e
\gamma)$ respectively, we have stronger bounds on their product thanks
to $Br(\mu\rightarrow e \gamma)$ experimental sensitivity.
These limits are important in order to get the largest amount of information 
on SUSY flavour symmetry breaking.\\
Summarizing the results found in this first stage, we can say that MI approximation produce 
the same features as the exact calculation even if strong cancellations occur.\\  
The approach works better in the $21$ sector, up to $(5\!-\!10)\%$ approximation level 
until $\delta_{21}\leq 0.2$, than in the 32 sector, where, large off
diagonal flavour conserving terms can induce a $(40\!-\!50)\%$ 
deviation from the full computation.\\
The GMI approximation works very well, as the MI approximation, 
except for some special regions  where induces a $(20\!-\!30)\%$
approximation with respect to the exact calculation and produces the cancellations
in a shift region respect to the exact case. In conclusion, we can say that, 
except for fine-tuned cases, the last approach is very satisfactory.\\ 
In a second stage, being our aim to find constraints in the $RR$
sector, we examined other LFV processes as
$\mu\rightarrow eee$ and $\mu\rightarrow e\ in\ Nuclei$.\\
We found that the last process suffers from a bigger cancellation problem than 
$\mu\rightarrow  e\gamma$ (in the $RR$ sector) while $\mu\rightarrow eee$ 
does not.\\ 
However, an interesting feature is that $\mu\rightarrow e\gamma$ and 
$\mu\rightarrow e\ in \ Nuclei$ amplitudes have cancellations in different regions, so, if 
we combine the two processes, we obtain a more stringent bound ($\delta_{21}^{RR}\leq$  0.2)
than the one coming from $\mu\rightarrow eee$ ($\delta_{21}^{RR}\leq$  0.4).  
It is noteworthy that the study of combined processes allows to extract additional
information respect to an individual analysis of all these processes.
In particular, it makes it possible to put bounds 
on sectors previously unconstrained by $\mu\rightarrow e \gamma$.\\ \\

\textbf{Acknowledgements}\\ I thank A. Masiero, R. Petronzio, N. Tantalo, S.K. Vempati and O. Vives
for useful discussions.
I also acknowledge the hospitality of the department of physics of Padova,
where part of this work was carried out. 
\appendix
\section{Loop functions}
In this appendix we report the explicit expressions for the loop functions appearing in the text:
$$f_{1n}(a)=\frac{-17a^3+9a^2+9a-1+6a^2(a+3)\ln a}{24(1-a)^5}$$
$$ f_{2n}(a)=\frac{-5a^2+4a+1+2a(a+2)\ln a}{4(1-a)^4}$$
$$f_{3n}(a)=\frac{1+2a\ln a-a^2}{2( 1-a)^3}$$
$$ f_{1c}(a)=\frac{-a^3-9a^2+9a+1+6a(a+1)\ln a}{6(1-a)^5}$$
$$ f_{2c}(a)=\frac{-a^2-4a+5+2(2a+1)\ln a}{2(1-a)^4}$$
$$ I_{1n}(a)=\frac{3a^4+44a^3-36a^2-12a+1-12a^2(2a+3)\ln a}{24(1-a)^6}$$
$$ I_{2n}(a)=\frac{a^3+9a^2-9a-1-6a(a+1)\ln a}{4(1-a)^5}$$
$$ I_{1c}(a)=\frac{10a^3+9a^2-18a-1-3a(3+6a+a^2)\ln a}{6(1-a)^6}$$
$$ I_{2c}(a)=\frac{3a^2-3-(a^2+4a+1)\ln a}{(1-a)^5}.$$

\end{document}